\documentclass[useAMS,usenatbib]{mn2e}
\usepackage{epsfig}
\usepackage{longtable}
\usepackage{times}

\def\mo{M$_\odot$}

\def\cm3{cm$^{-3}$}
\def\kms{km~s$^{-1}$}

\def\lsun{L$_{\odot}$}

\def\etal{et al.}

\def\beq{\begin{equation}}
\def\eeq{\end{equation}}

\def\sles{\lower2pt\hbox{$\buildrel {\scriptstyle <}
   \over {\scriptstyle\sim}$}}
\def\sgreat{\lower2pt\hbox{$\buildrel {\scriptstyle >}
   \over {\scriptstyle\sim}$}}

\def\aj{AJ}
\def\apj{ApJ}
\def\apjl{ApJL}
\def\apjs{ApJS}
\def\aap{A\&A}
\def\araa{ARA\&A}
\def\aaps{A\&AS}
\def\mnras{MNRAS}
\def\nat{Nature}
\def\pra{Phys.~Rev.~A}
\title[SN 1994W:  An Interacting Supernova or Two Interacting Shells?]{SN 1994W:
An Interacting Supernova or Two Interacting Shells?}

\author[Luc Dessart, D. John Hillier, Suvi Gezari, St\'ephane Basa, and Tom Matheson]
{Luc Dessart$^{1,2}$\thanks{E-mail: luc@as.arizona.edu}, D. John Hillier$^3$,
Suvi Gezari$^4$, St\'ephane Basa$^5$, and Tom Matheson$^6$\\
$^1$ Department of Astronomy and Steward Observatory,
                 The University of Arizona, Tucson, AZ 85721\\
$^2$ Department of Astrophysical Sciences, Princeton University, Princeton, NJ 08544\\
$^3$ Department of Physics and Astronomy, University of Pittsburgh, PIttsburgh, PA 15260 \\
$^4$ California Institute of Technology, Pasadena, CA 91125 \\
$^5$ Laboratoire d'Astrophysique de Marseille, France \\
$^6$  NOAO Gemini Science Center, 950 North Cherry Avenue, Tucson, AZ 85719
}

\begin{document}

\date{Accepted . Received }

\pagerange{\pageref{firstpage}--\pageref{lastpage}} \pubyear{2007}

\maketitle

\label{firstpage}

\begin{abstract}

We present a multi-epoch quantitative spectroscopic analysis of the Type IIn SN 1994W, an event
interpreted by Chugai et al. as stemming from the interaction between the ejecta of a
SN and a 0.4\,\mo\ circumstellar shell ejected 1.5\,yr before core collapse.
During the brightening phase, our models suggest that the source of optical radiation
is not unique, perhaps associated with an inner optically-thick Cold Dense Shell (CDS) and
outer optically-thin shocked material. During the fading phase, our models support a {\it single}
source of radiation, an hydrogen-rich optically-thick
layer with a near-constant temperature of $\sim$7000\,K that recedes from a radius
of 4.3$\times$10$^{15}$ at peak to 2.3$\times$10$^{15}$\,cm 40 days later.
We reproduce the hybrid narrow-core broad-wing line profile shapes of SN 1994W
at all times, invoking an optically-thick photosphere exclusively (i.e., without any
external optically-thick shell). In SN 1994W, slow expansion makes scattering with thermal electrons
a key escape mechanism for photons  trapped in optically-thick line cores, and allows the resulting
broad incoherent electron-scattering wings to be seen around narrow line cores.
In SNe with larger expansion velocities, the thermal broadening due to incoherent
scattering is masked by the broad profile and the dominant frequency redshift
occasioned by bulk motions. Given the absence of broad lines at all times and the very
low $^{56}$Ni yields, we speculate whether SN 1994W could have resulted from an interaction
between two ejected shells without core collapse. The high conversion efficiency of kinetic
to thermal energy may not require a SN-like energy budget for SN1994W.


\end{abstract}

\begin{keywords}
radiative transfer -- stars: atmospheres -- stars:
supernovae: individual: 1994W
\end{keywords}

\section{Introduction}

  Understanding the diversity of supernova (SN) spectra and light curves is a considerable challenge.
Although some SN classes suffer less and less ambiguity, as in the thermonuclear incineration
of a Chandrasekhar-mass white dwarf leading to a Type Ia SN, or as in the collapse of the degenerate
Chandrasekhar-mass Fe or O/Ne/Mg core of a massive star leading to Type II or Type Ib/c SNe,
a unified picture of stellar explosions
is compromised by the growing number of atypical SNe that emerge from deeper and more frequent
searches for transient phenomena in the local and distant Universe. One such type of peculiar SNe
are ``n''-suffixed (i.e. IIn, Ibn, Ian), in reference to their atypical narrow line profiles.
This suggests that despite the bright,
supernova-like, visual display, the expansion rate of the radiating layer is small, i.e. of a few 100\,\kms,
in contrast with the large ejecta velocities, of a few 1000\,\kms, associated with SN explosions.
Paradoxically, these events can boast a huge bolometric luminosity --- a few 10$^9$\,\lsun\,
sustained for weeks (as in SN 1994W) but associated with a spectral energy distribution (SED)
typical of a 10,000\,K blackbody, hence implying ``photospheric'' radii of a few 10$^{15}$\,cm
($ L \sim 2 \times 10^9 (R_{15})^2 (T_4)^4 L_{\odot}$, where $R_{15}$ is the radius in units of
10$^{15}$\,cm and $T_4$ the temperature in units of 10$^4$\,K).

All SN ejecta yield photospheric radii of 10$^{15}$\,cm a few weeks after explosion, but the
remarkable property of  Type IIn SNe is that they achieve this with an apparent slow expansion.
The common interpretation is that Type IIn SNe interact with material present in the direct
environment of the exploding star, forming the ``photosphere'' where that external material resides,
at a few 10$^{15}$\,cm. If its mass is on the order of the ejecta mass, this outer material
can cause a significant deceleration  of the ejecta, with an efficient conversion of kinetic
energy into internal and radiative energy.

The narrow line cores are accompanied, in a subset of Type IIn SNe, by broad (and symmetric) line
wings which are generally interpreted as arising from multiple electron-scattering of line
photons in the circumstellar (CS) shell with which the ejecta collide (Chugai 2001).
These optical ``peculiarities''
are also often associated with the detection of X-ray and/or radio emission, as well as significant
spectral variations.  Although interaction is evident in many SNe (see, e.g.,
Chevalier \& Fransson 1994, 2001; Fransson et al. 1996), Type IIn SNe are objects
in which the interaction is central,  and perhaps ``makes'' the SN (i.e., the interaction
can increase the luminosity so that it is easier to detect).

  Given the rich mass loss history and the many alternate channels of evolution, massive stars
are the prime candidates for such interactions, in particular the hydrogen-rich, Type II, SNe.
A few well-documented examples that reveal the wealth and the diversity of phenomena hosted by such
IIn events are SN 1988Z (Stathakis \& Sadler 1991; van Dyk et al. 1993; Turatto et al. 1993;
Chugai \& Danziger 1994; Fabian \& Terlevich 1996; Aretzaga et al. 1999; Williams et al. 2002;
Schlegel \& Petre 2006), SN 1994W (Sollerman et al. 1998, hereafter SCL; Tsvetkov 1995; Schlegel 1999;
Chugai et al. 2004a, hereafter C04), SN 1995G (Pastorello et al. 2002; Chugai \& Danziger 2003),
SN 1995N (Fox et al. 2000; Fransson et al. 2002; Mucciarelli et al. 2006;
Chandra et al. 2005; Zampieri et al. 2005), SN 1997eg (Salamanca et al. 2002; Hoffman et al. 2007),
SN 1998S (Bowen et al. 2000; Gerardy et al. 2000; Leonard et al. 2000; Liu et al. 2000;
Anupama et al. 2001; Chugai 2001; Lentz et al. 2001; Fassia et al. 2000,2001;
Chugai et al. 2002; Pooley et al. 2002; Fransson et al. 2005; Pozzo et al. 2004,2005).
Despite the common IIn SN type, the above sample of SNe do not boast a uniform set of properties,
some being bright for weeks, some for months or even years; some showing a clear brightness
plateau after peak, others fading by a few magnitudes in just a few weeks, etc.
Objects classified as ``n-''type SNe are in fact shaking the common understanding of what makes a SN.

The Type IIn SN 2006gy, the most luminous SN ever seen, may have originated from a
pair-instability explosion (Smith et al. 2007; Ofek et al. 2007; Smith \& McCray 2007),
or consecutive pair-instability pulsations (Woosley et al. 2007) in a super-massive star.
Also recently, SN 2006jc joined SN 1999cq and SN 2002ao to consolidate the rare SN Ibn type, characterized
by narrow lines of helium instead of hydrogen, an interpretation that conflicts with the notion that pre-SN
mass ejections are generally associated with hydrogen-rich stars (e.g., Luminous Blue Variables, LBVs,
see Davidson \& Humphreys 1997; Matheson et al. 2000; Foley et al. 2007; Pastorello et al. 2007).
Interaction with CS material has also been invoked in the Type Ia SN 2002ic following
the observations of narrow hydrogen-line emission in the optical spectrum
(Hamuy et al. 2003; Chugai \& Yungelson 2004; Chugai et al. 2004b; Deng et al. 2004;
Kotak et al. 2004; Wang et al. 2004; Wood-Vasey et al. 2004; Benetti et al. 2006; Han \& Podsialowski 2006;
Wood-Vasey \& Sokoloski 2006; Chugai \& Chevalier 2007).

Modeling such an interaction is a complicated radiation hydrodynamics problem, in which
the uncertainties in the properties of both shells prior to interaction
add to those of the interaction itself. The diversity of observations,
suggests that a similar diversity is to be found in the circumstances of the interaction.
Focusing from now on on the massive star progeny, the outer, CS, material may arise
from the pre-SN steady-state wind mass loss, an LBV outburst,
perhaps a violent pair-instability pulsation, or some yet unidentified ejection mechanism
occurring recurrently (and without core collapse)
or immediately prior to core-collapse. The inner shell that rams into this CS shell
may arise from a core collapse SN explosion (which may be neutrino-, acoustic-,
or magnetically-driven; see Woosley \& Janka 2005 for a review),
a pair-instability pulsation, a pair-instability explosion, or some other form of explosion.
For a SN-like display to occur, the inner shell must be fast to catch up with the CS material
and cause a violent shock. A short delay between the two mass
ejections will raise the probability for detection. A large kinetic energy is involved, and the conversion
of kinetic to thermal energy supplies the internal/radiant energy inferred from the bolometric light curve.
Moreover, mass/energy distribution of both shells may not be spherical (Leonard et al. 2000; Hoffman et al. 2007)
and thus, viewing effects may complicate further the dynamics of the interaction and the interpretation of the
emitted light. Finally, the bolometric display may be altered by an additional contribution
from radioactive decay of unstable isotopes, which are normally associated with SN ejecta.


An attempt to quantitatively interpret the Type IIn SN 1994W was undertaken by
C04 who performed radiation-hydrodynamics and radiative-transfer calculations
to model very high quality multi-epoch spectroscopic and photometric observations (see also
SCL). For SN 1994W, C04 associate the narrow line core with
an expanding CS envelope. The broad line wings arise from a combination of shocked cool gas in
the forward post-shock region, and multiple electron scattering in the CS envelope. They associate
the absence of
broad P~Cygni line profiles with obscuration by an optically-thick cold dense shell (CDS)
that forms at the interface of SN ejecta and a CS envelope. They infer a CS envelope with
a particle density $n \sim 10^9$\,cm$^{-3}$, a radial extent of a few 10$^{15}$\,cm,
an electron-scattering optical depth of $\sgreat$2.5, explosively ejected $\sim$1.5\,yr prior to the
SN explosion. The light curve shows a rise time to peak of about 30 days (after the reference date
1994 July 14), followed by a slow decline by about two magnitudes until 110\,days,
and a sudden decline beyond that date (see Fig.~1 in SCL).

Despite all the complexities that surround this ejecta/CS-envelope interaction,  the
inferred presence by C04 of an optically-thick CDS suggests that the approach we follow
for the modeling of photospheric-phase Type II-Plateau (II-P) SNe (Dessart \& Hillier 2005ab, 2006, 2008;
Dessart et al. 2008) may also apply here. In this work, we provide insights into the
spectroscopic and light-curve evolutions of SN 1994W between day 20 and 100, thus
covering from $\sim$10 days before peak, through to the slow decline that follows
until 10 days before the steep brightness drop.

The main results from this work are that although the outward-moving CDS is likely opaque until
the peak of the light curve, the photosphere recedes both in mass and radius for post-peak
times as the material cools and recombines. In this context, the sharp drop at 110 days is
the transition to the nebular phase, when the  ejecta/CSM is entirely optically-thin,
analogous to the end of the plateau phase in Type II-P SNe. We also
find that the broad wings on the Balmer lines can be explained by multiple electron scattering
in the photosphere --- we do not need multiple density structures to explain the observed spectrum.


In the next section, we discuss the reddening and distance used in our study of SN 1994W.
In Section \ref{sect_time}, we evaluate various timescales that characterize Type IIn SNe. We then
summarize our modeling approach in Section \ref{sect_model}, presenting the various approximations
we make to mimic as best we can the complicated configuration of the interacting SN 1994W.
In Section \ref{sect_results}, we present the ejecta properties we infer from the modeling of
the photometric and spectroscopic observations at multiple epochs during its optically-bright phase.
We reproduce the narrow line core and the broad line wings by invoking a single
region of emission/absorption (the photosphere), rather than the CDS and CS-envelope configuration.
In Section \ref{sect_es}, we discuss the line formation process in our models
of SN 1994W, and emphasize the important role of electron-scattering in some Type IIn SN photospheres
that are characterized by low, and perhaps nearly constant, expansion velocities. We finally present
our conclusions and discuss the implication of our results for interacting SNe in a more general
context in Section \ref{sect_discussion}.

%
%
%
%
%
%

\section{Reddening \& Distance\label{sect_red_dist}}

Following SCL we adopt a distance to SN 1994W of 25.4\,Mpc which is based on
$H_0=65$\,\kms\,Mpc$^{-1}$ and the Virgo infall model of Kraan-Korteweg (1986). This is somewhat larger
than the distance estimate of 18.0\,Mpc by Gao \& Solomon (2004) which used
$H_0=75$\,\kms\,Mpc$^{-1}$ and was corrected for motion of the local group. While the choice of distance
will affect the adopted luminosity and inferred photospheric radii, other conclusions in this paper
(including $T_{\rm phot}$) are not affected.
The reddening is more problematical since it can affect the relative flux distribution.
SCL find  $E(B-V)=0.17 \pm 0.06$\,mag based on the equivalent width of the Na{\sc i}D doublet and its correlation
with $E(B-V)$ while C04 estimate 0.15\,mag from blackbody fits to spectra at
several different dates. As the galactic extinction in the direction of SN 1994W is low,
most of the reddening must be internal to the host galaxy (SCL). Our present model, particularly
of spectra past the peak, supports the SCL reddening although values of  0.15 to 0.20\,mag are also
compatible with the observations. In this work, we employ a reddening of 0.15-0.17, but also explore
the effects of reducing the reddening to 0.07 on day 21.

\section{Timescales}
\label{sect_time}

A typical expansion velocity for SN 1994W, as measured from the narrow
line profiles, is ~800\,\kms. We can use this to define an expansion timescale
using a typical photospheric radius of $4 \times 10^{15}$\,cm. This gives
an expansion timescale of 1.6 years. As noted earlier, this, together with the large luminosity,
and the difficulty of creating SNe with a low expansion velocity,
are reasons for the argument that the 800\,\kms\, is not an intrinsic SN
expansion velocity. C04 suggest it is the velocity of CS material that was ejected prior to the SN explosion.
Alternatively, it could reflect the velocity of a shell of gas
that has arisen from the interaction of circumstellar gas with the SN ejecta.
The largest red-supergiants have radii of $~10^{14}$ cm
(see, e.g., Levesque et al. 2005), significantly smaller than the inferred photospheric radius.

The light travel time across the SN is $2R/c \sim 3.1$ days. This is marginally
significant, and indicates that the observed light curve will be averaged
over this timescale. The diffusion time is of order $\tau R/cn$, where $n$ is the exponent
of the power law density. This will also have a significant impact on the observed light curve,
although it is less clear whether it will affect our spectroscopic modeling in which we
fix the luminosity at the base of the photosphere.

\begin{table*}
  \centering
  \begin{minipage}{170mm}
  \caption{Model Characteristics for SN 1994W.
For each date in our sample of observations, we provide the following
CMFGEN model parameters: Base comoving-frame luminosity $L_{{\rm CMF},R_0}$ and
emergent observer-frame luminosity $L_{{\rm OBS},R_{\rm Max}}$ (in 10$^8$\,\lsun),
photospheric conditions describing
the electron temperature $T_{\rm phot}$ (in K), the radius $R_{\rm phot}$ (in 10$^{15}$\,cm),
the velocity $V_{\rm phot}$ (in \kms), the mass density (in 10$^{-14}$\,g\,cm$^{-3}$), and
the free-electron density $N_{\rm e, phot}$ (in 10$^9$\,cm$^{-3}$).
In all models, a density exponent $n=10$ characterizes the density law
$\rho(R) = \rho_{\rm phot} (R_{\rm phot}/R)^n$.
$M_V$ and $m_V$ correspond to $V$-band absolute, and reddened plus distance-diluted (using a distance of
25.4\,Mpc and a reddening of 0.17 with $R_V=3.1$), synthetic magnitudes.
Because there are clear inconsistencies between our model and the observations on days 21 and 31,
we quote the corresponding model parameters on those dates only for completeness -
we do not suggest they accurately describe SN 1994W on these two dates.
$^a$: Days after 14.0 July 1994.
$^b$: So-called hot model, used to fit observations on the 4th and the 14th of August, both poorly.
$^c$: So-called cool model, used to fit observations on the 4th and on the 14th of August (but poorly
for both dates), and on September 1 (satisfactorily).
\label{tab_model_94W}
}
  \begin{tabular}{lcccrcccccc}
  \hline
Day & Phase$^a$ & $L_{{\rm CMF},R_0}$ & $L_{{\rm OBS},R_{\rm Max}}$& $T_{\rm phot}$ &
$R_{\rm phot}$ & $V_{\rm phot}$ & $\rho_{\rm phot}$ & $N_{\rm e, phot}$ & $M_V$ & $m_V$  \\ 
YY--MM--DD & Days & \multicolumn{2}{c}{(10$^8$ $L_{\odot}$)}& (K)& (10$^{15}$\,cm) &(\kms) &
(10$^{-14}$\,g\,cm$^{-3}$) & (10$^9$\,cm$^{-3}$) & \multicolumn{2}{c}{synthetic}  \\
\hline
1994-08-04$^b$ &  21.5   &    66.0   &     61.7  & 10350  &     2.86  &   830  &   0.82 & 3.3  & -18.78  & 13.37  \\ 
1994-08-14$^b$ &  31.5   &    66.0   &     61.7  & 10350  &     2.86  &   830  &   0.82 & 3.3  & -18.78  & 13.37  \\ 
1994-09-01$^c$ &  49.5   &    47.0   &     42.0  &  7480  &     4.32  &   830  &   0.62 & 2.1  & -18.80  & 13.75 \\ 
1994-09-09     &  56.9   &    27.0   &     24.0  &  7450  &     3.33  &   840  &   0.84 & 2.7  & -18.37  & 14.18 \\ 
1994-10-01     &  79.5   &    9.8    &     8.7   &  6310  &     2.37  &   790  &   2.7  & 5.2  & -17.61  & 14.94 \\ 
1994-10-11     &  89.5   &    9.0    &     8.0   &  6020  &     2.32  &   712  &   7.6  & 7.5  & -17.59  & 14.96 \\ 
\hline
\hline
\end{tabular}
\end{minipage}
\end{table*}

\section{Model Presentation}
\label{sect_model}



  In hydrogen-rich environments, having an optically-thick layer simplifies considerably the
radiative transfer problem. The associated large bound-free opacities ensure that the
radiation is thermalized before escaping through the photosphere ---
all photons at depth will be absorbed and re-emitted according to a blackbody distribution
characterized by the local electron temperature, irrespective of the details of the interaction.
Hence, we have a setup that is analogous to that of a typical stellar atmosphere.
Not accounted for in this configuration is the potential contribution
from the optically-thin layers above the photosphere that have been
shocked. These regions can contribute to the electromagnetic display
directly, as they often do in the X-ray and radio ranges, or alter the photospheric conditions
through external irradiation. How well we reproduce the observations may be a gauge on how much
these regions contribute to the total SN luminosity and affect the photospheric conditions.
In the case of SN 1994W we can reproduce the observations after the peak by photospheric
radiation alone.  However before the peak we cannot fully reproduce the observations accurately,
possibly indicating that optically-thin gas makes an important contribution. Alternatively,
neglected effects such as departures from spherical symmetry and time dependent effects,
may be important at such early times (Dessart \& Hillier 2008).



We model the observations with the non-LTE steady-state one-dimensional radiative transfer code CMFGEN
(Hillier \& Miller 1998; Dessart \& Hillier 2005a) which solves self-consistently the radiative transfer
equation and the statistical equilibrium equations under the constraint of radiative equilibrium.
Of particular relevance for this work, CMFGEN treats accurately the electron-scattering source function.
With our adopted distance and a reddening of $E(B-V)=0.17$\,mag a blackbody emitter, with a temperature
of 10$^4$\,K and a radius of 10$^{15}$\,cm, would have an observed V-band magnitude of $\sim$14.5,
in close agreement with the observed value around day 50 (see Fig.~1 of SCL98).
We therefore started our model analysis with a typical Type II-P SN model just prior to hydrogen
recombination (as in Dessart \& Hillier 2005a,2006 or Dessart et al. 2008), and modified the radius
and luminosity to agree with the above brightness estimate. We also adopt a supergiant-like composition,
i.e. $X_{\rm H}=$0.55, $X_{\rm He}=$0.44, $X_{\rm C}=$5.3$\times$10$^{-4}$,
$X_{\rm N}=$2$\times$10$^{-3}$, $X_{\rm O}=$2.8$\times$10$^{-3}$. Metal abundances are taken
at the solar value. We use the same model atom for all investigations presented here, with
H{\sc i}, He{\sc i}, C{\sc i}, C{\sc ii}, N{\sc i}, N{\sc ii}, Na{\sc i}, Mg{\sc ii}, Si{\sc ii},
Ca{\sc ii}, Al{\sc ii}, Al{\sc iii}, O{\sc i}, O{\sc ii}, S{\sc ii}, S{\sc iii}, Cr{\sc ii}, Cr{\sc iii},
Mn{\sc ii}, Mn{\sc iii}, Ti{\sc ii}, Ti{\sc iii}, Co{\sc ii}, Co{\sc iii}, Ni{\sc ii},
Ni{\sc iii}, Fe{\sc ii}, Fe{\sc iii}, Fe{\sc iv} (details on the levels treated are omitted, but our
choice is such that increasing the number of levels does not alter the computed ejecta properties nor
the emergent synthetic spectrum).

 The atomic data come from a wide variety of sources, the  Opacity Project (Seaton 1987;
The Opacity Project Team 1995, 1997), the Iron Project (Pradhan et al. 1996; Hummer et al. 1993),
Kurucz (1995), and the Atomic Spectra Database at NIST Physical Laboratory being the principal sources.
Much  of the Kurucz data was obtained directly from the Center for Astrophsyics (Kurucz 1988, 2002).
Individual sources of atomic data include the following: Bautista \& Pradhan (1997),
Becker \& Butler (1995), Butler et al. (1993), Fuhr et al. (1988), Kingdon \& Ferland (1996),
Luo \& Pradhan (1989), Luo et al. (1989), Mendoza (1983), Mendoza et al. (1995), Nahar (1995, 1996),
Nahar \& Pradhan (1996), Neufeld \& Dalgarno (1987), Nussbaumer \& Storey (1983, 1984),
Peach et al. (1988), Storey (1988), Tully et al. (1990), Wiese et al. (1966), Wiese et al.
(1969), Zhang \& Pradhan (1995, 1997).


Dynamical simulations by Chevalier (1982) and C04 suggest that in the present
context a very steep density fall-off would prevail above the CDS, while the velocity would be
decreasing from the CDS outward into the shocked CS envelope.
Adopting a power-law density distribution
of the form $\rho(R)=\rho_0 (R_0/R)^n$, where $\rho_0$ and $R_0$ are, respectively, the density
and the radius at the optically-thick model base, we enforce a steep density fall off by taking a
density exponent $n$ equal to 10. Flatter density distributions, as in a wind solution with $n=2$,
yield optical line fluxes that are systematically stronger than observed. Steeper density
distributions lead to weaker line fluxes and/or absorption lines. With our current
approach and from extensive experimentation, we indeed find that the density distribution
at the photosphere has to be steep, in fact comparable to that used for the modeling of
Type II-P SNe.\footnote{Note that recombination to a neutral state may attenuate
the spectral dependence on the density distribution above the photosphere.}

\begin{figure*}
\epsfig{file=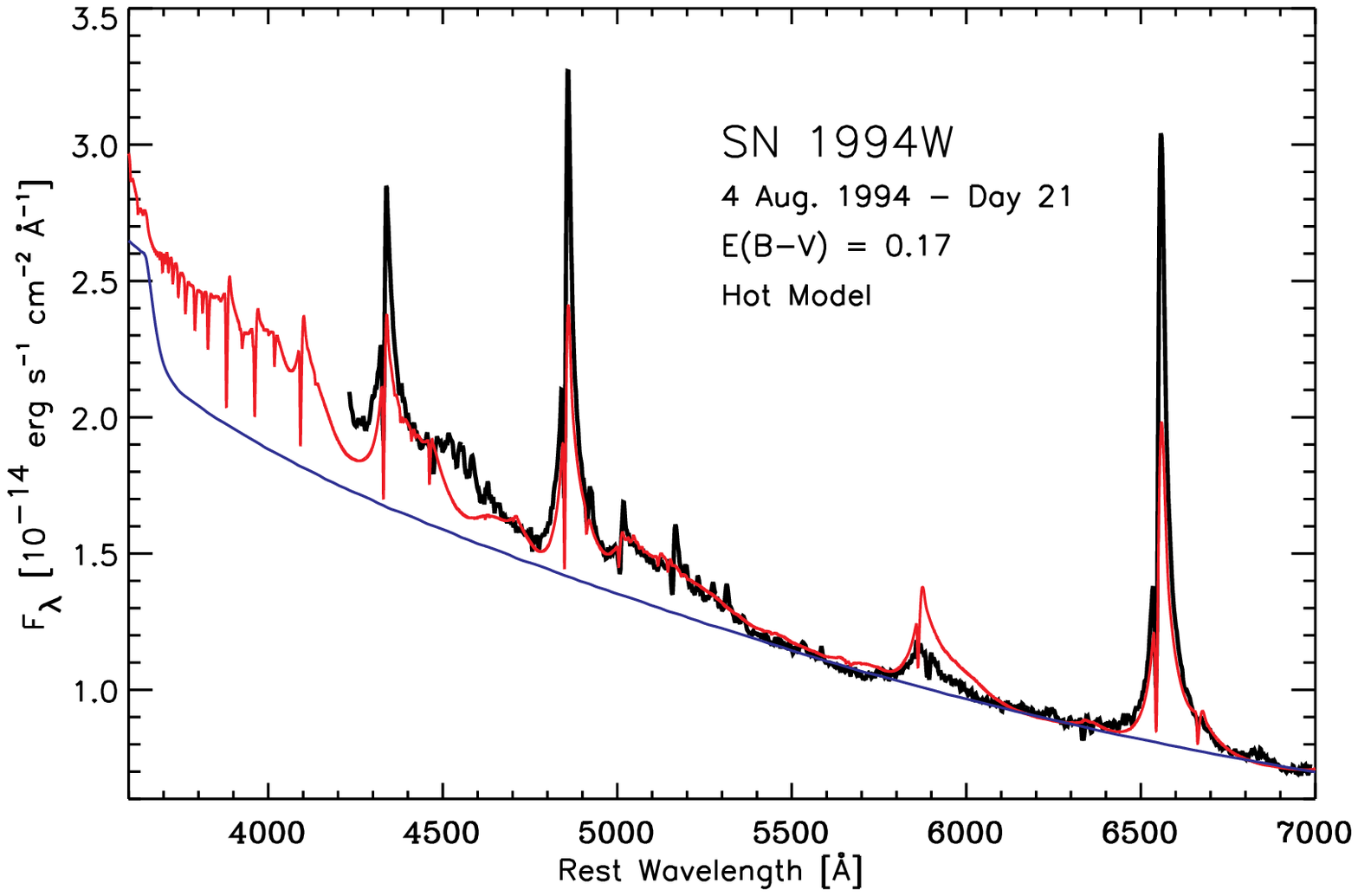,width=5.5cm}
\epsfig{file=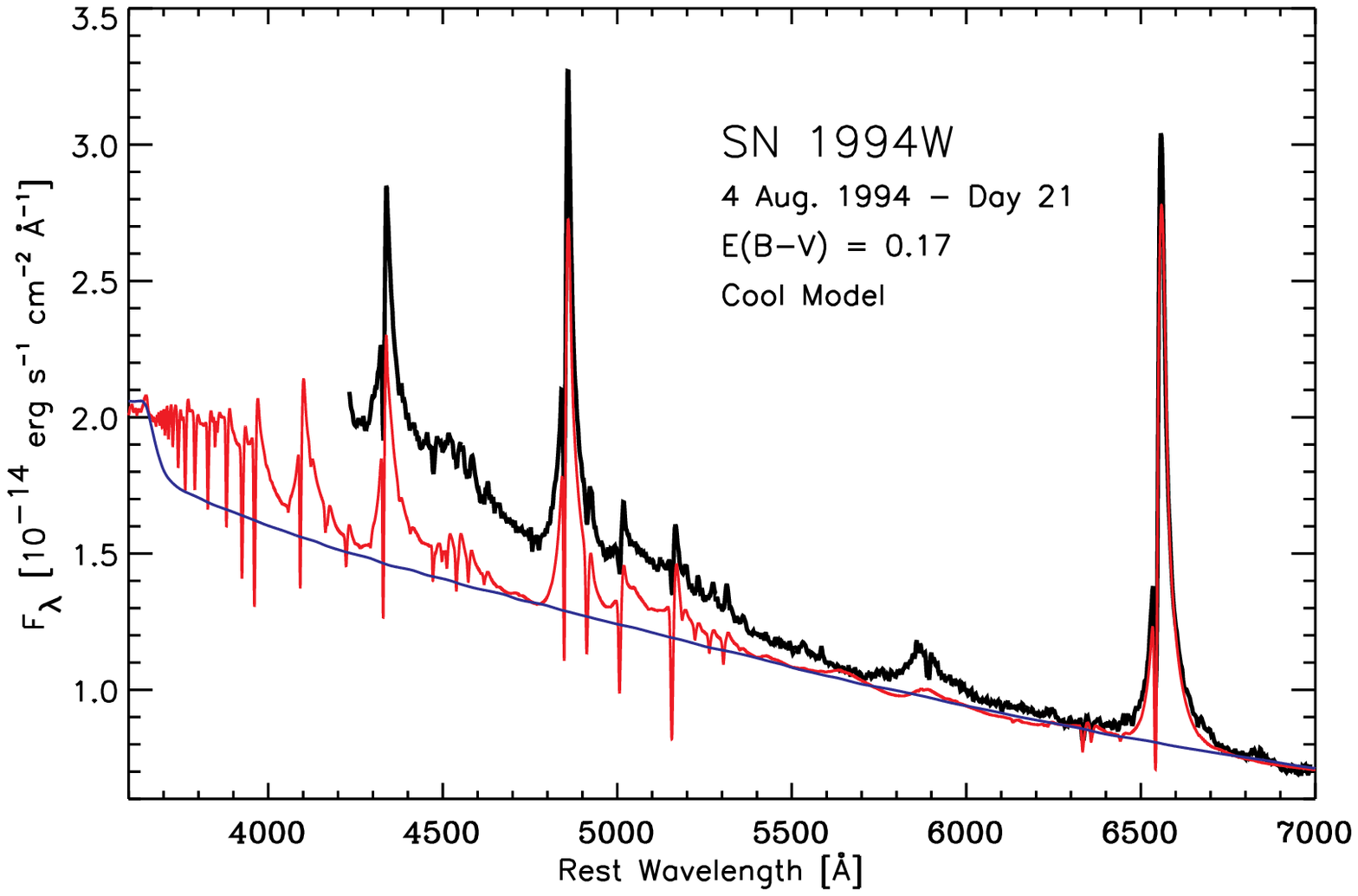,width=5.5cm}
\epsfig{file=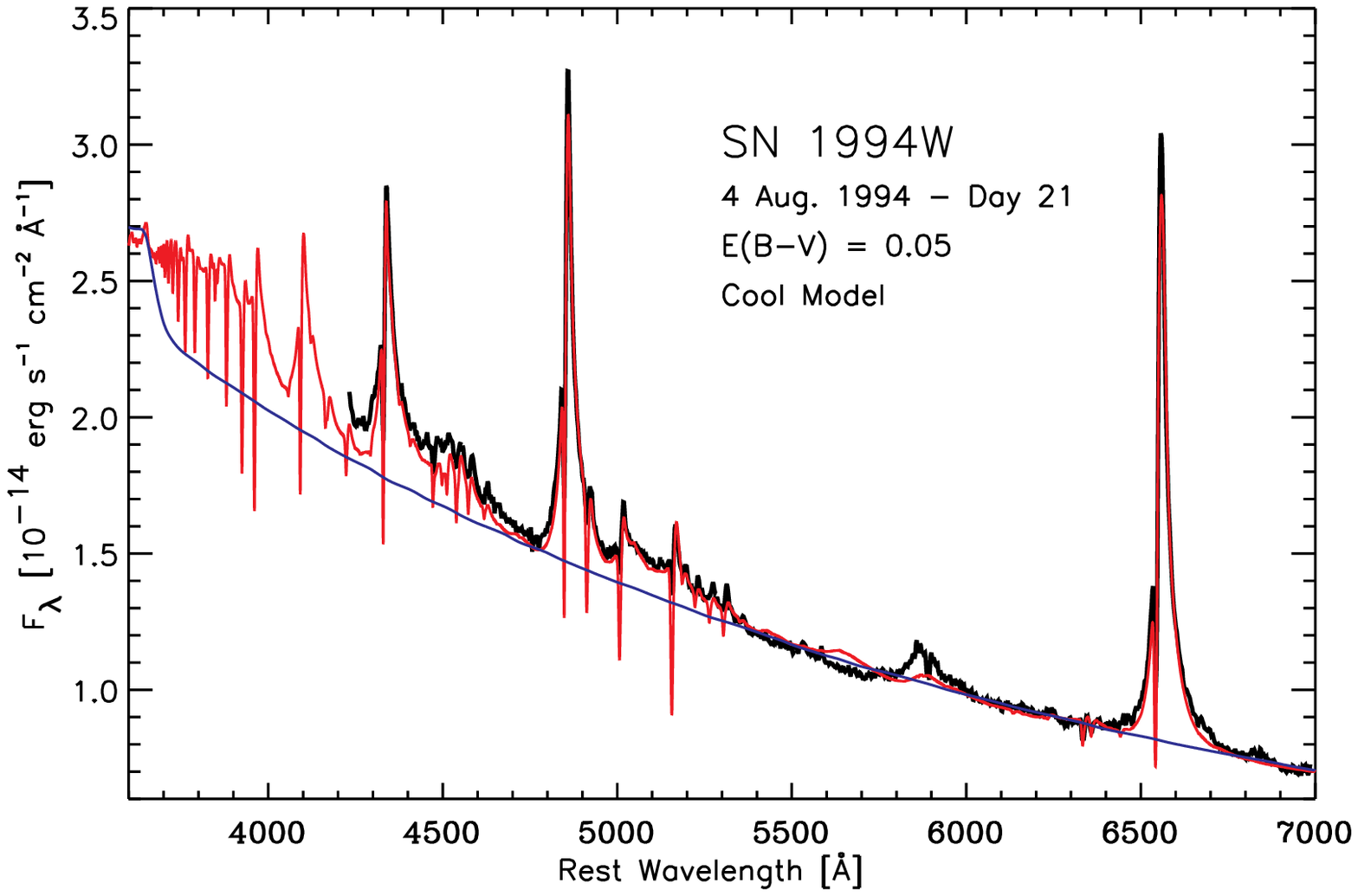,width=5.5cm}
\caption{{\it Left:} Comparison between the reddened (E(B-V)=0.17\,mag) full (red) and continuum-only (blue)
synthetic spectrum and the observations of SN 1994W on the 4th of August 1994 (day 21; black).
The synthetic flux is scaled by a factor of 1.06 to adjust to the absolute level
of the observed flux.
The ejecta ionization predicted with our model parameters reproduces the slope of the SED
adequately, but it over-estimates the strength of He{\sc i}\,5875\,\AA\ and underestimates
the strength of Fe{\sc ii} lines. Note however the good fit to the Balmer lines.
{\it Middle:} Same as left, but now with a cooler model used to fit observations
on 1994 September 1 (see Fig.~\ref{fig_94W_0901}; we use a flux scaling of 0.93).
Note the bad fit to the SED slope, but the improved fit to observed line profiles.
{\it Right:} Same as middle, but using a reddening of 0.05 instead of 0.17.
Now, the fit to both the continuum and lines is good. What causes this combination
of blue continuum and lines of low-ionization species is unclear. Being short-lived,
it may be related to shocked material above the photosphere, rather than to large
changes in photospheric conditions.
Reducing the reddening is only a proxy for getting a good fit, for exploratory purposes,
to the data taken on this date.\label{fig_94W_0804}
}
\end{figure*}

The transfer equation solver in CMFGEN
requires {\it monotonically} expanding ejecta. In SN ejecta, the steep density profile
tends to make the line and continuum formation quite confined in space, so the material
properties on a large scale tend to matter little compared to the properties in the
immediate vicinity of the photosphere. In our approach, we assume homologous expansion for
simplicity, and adjust the ejecta velocity to match the observed line profile widths.
Note that in the line transfer problem, it is the magnitude of the velocity gradient
at the photosphere, rather than its sign, that matters; (see Sobolev 1960, Castor et~al. 1975).
Obviously, this is a numerical convenience and we do not suggest that in reality the velocity increases linearly
with radius outwards, although it might. Later on, to test the dependence on the velocity gradient
at the photosphere, we try out different values for the exponent $\beta$ entering our parameterized velocity law,
i.e. $V(R) = V_0 (R/R_0)^{\beta}$, with $\beta$ varying from 1 (homologous expansion) to 0.2,
and 0.01 (near-constant velocity ejecta; here, $V_0$ is the model-base velocity).
Moreover, to ensure the radiation is thermalized
at the inner boundary, we extend our grid inwards to a radius where the Rosseland optical depth is $\sim$100.
Finally, when comparing to observations, we first redden our synthetic spectra with the
Cardelli law (Cardelli et al. 1989), and then adjust the synthetic flux (at most by a few percent)
to get the desired overlap. This is a convenience, equivalent to a change in photospheric radius or distance
(scaling with the square root of that scaling factor), that saves us from re-running a model that would
otherwise have the same properties and spectrum (see discussion in Dessart \& Hillier 2005a).

\section{Results}
\label{sect_results}

   We have performed a spectroscopic analysis of the data presented in SCL and C04 for the
observations taken on 1994 August  4 and 14 (Section \ref{sect_pre_peak}), September 1 and 9,
and October 1 and 11 (the last four dates are presented in
Section \ref{sect_post_peak}), which correspond to days 21, 31, 49, 57, 79, and 89 after the
14th of July 1994 (we adopt that reference date for compatibility with SCL, who associate
it with the ``optical outburst'' of SN 1994W).

These observations correspond to a phase of optical brightening
prior to day 30, followed by an 80-day long phase of slow optical fading
($\Delta m \sim 1.5$\,mag), eventually followed by a sudden fading of the SN past day 110.
We omit from our sample the lower quality data obtained on day 18,
as well as the late time, nebular spectrum obtained on day 121.
 Model parameters for each epoch are stored in Table~1, although we stress that the results
for days 21 and 31 are uncertain, and are given for completeness only.

All observed spectra have been de-redshifted assuming a heliocentric velocity of 1249\,\kms (C04).
Using the Cardelli et al. (1989) law, we redden our synthetic spectra to match observations,
using $E(B-V)=$0.17\,mag, unless otherwise stated.
We also provide, as online material, the list of all lines
between 3200\,\AA\ and 10500\,\AA\ that contribute an equivalent width (approximate since
computed using the Sobolev approximation, and quoted merely to indicate the main contributors;
Sobolev 1960) of at least 1\,\AA\ (in absolute value).
Each line, in absorption, emission, or both, may appear as a single feature, or may overlap
with neighboring lines to yield a complicated feature.

\begin{figure}
\epsfig{file=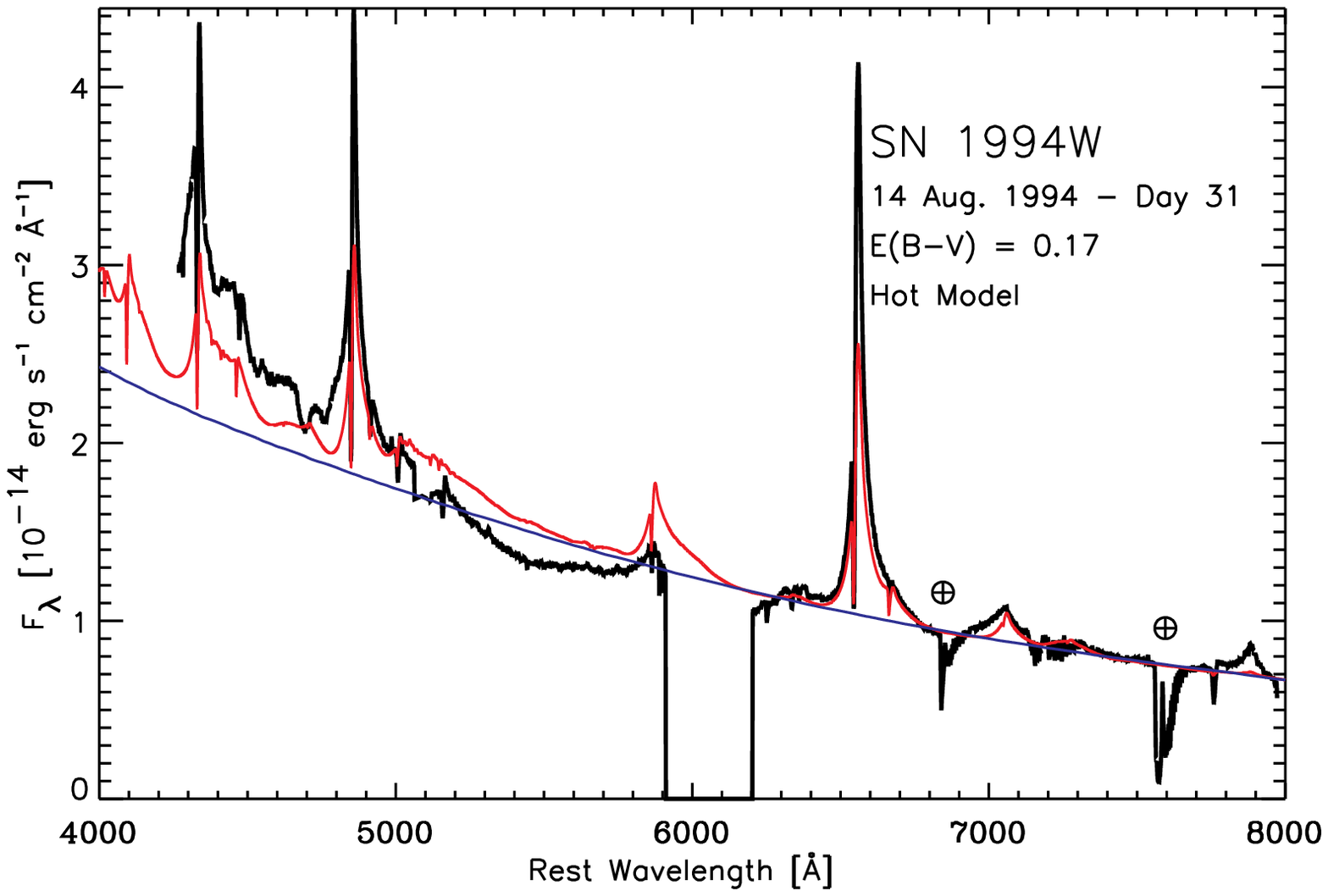,width=9cm}
\epsfig{file=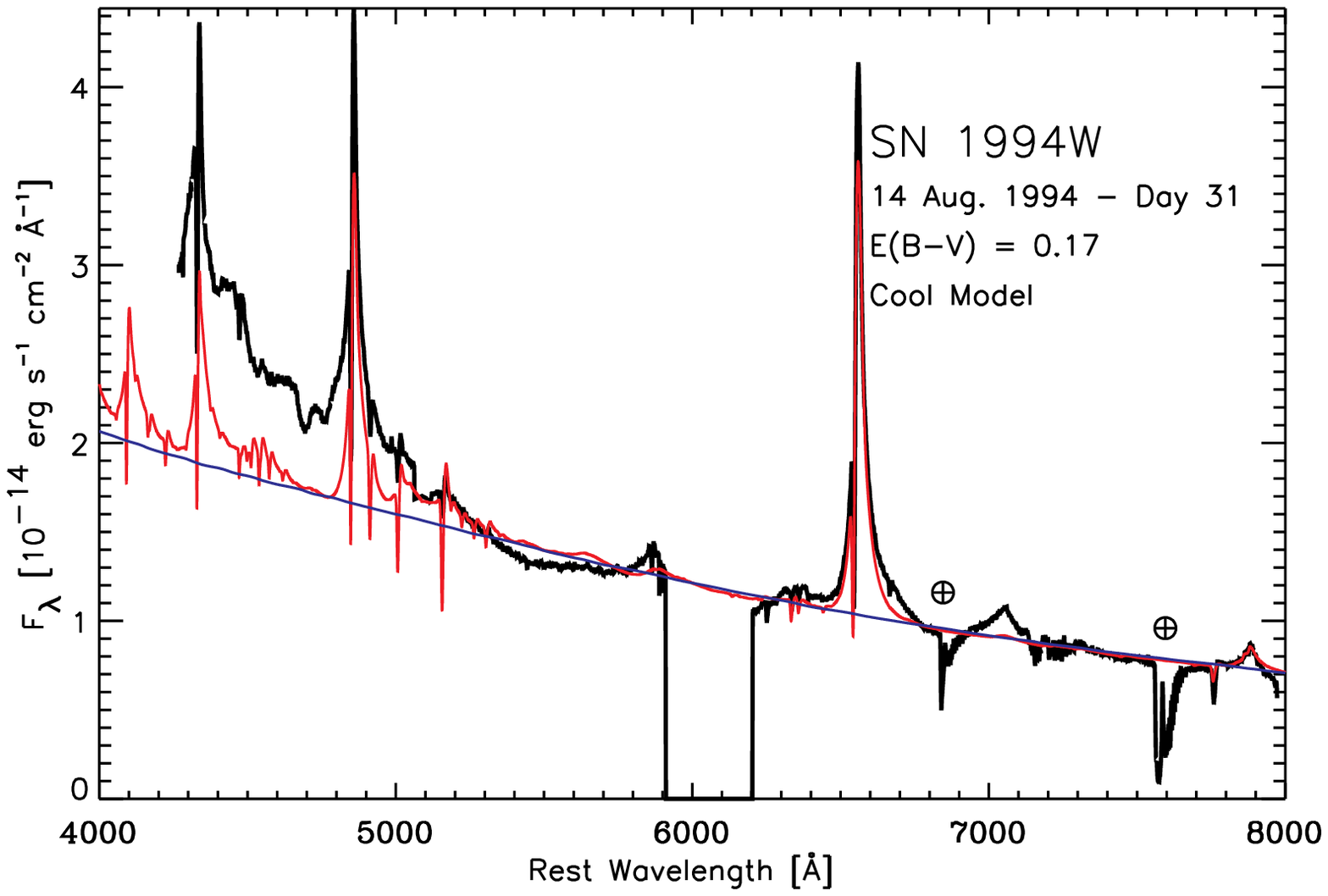,width=9cm}
\caption{{\it Top:} Comparison between the reddened (E(B-V)=0.17\,mag) full (red) and continuum-only (blue)
synthetic spectra for a hot model and the observations of SN 1994W on the 14th of August 1994 (day 31; black).
The synthetic flux is scaled by a factor of 1.37 to adjust to the absolute level
of the observed flux.
{\it Bottom:} Same as left, but now with the (cooler) model employed for the observations
of September 1 (we apply a flux scaling of 1.2).
Notice how the Mg{\sc ii}\,7877--7896\,\AA\ (He{\sc i}\,7065\,\AA) line is well fitted with
the cool (hot) model, but not both at the same time, suggesting two different sites/mechanisms
for the origin of these lines.
By contrast, the electron scattering wings are well reproduced in both models, despite
the ejecta ionization changes.
}\label{fig_94W_0814}
\end{figure}

  \subsection{Observations pre-peak optical brightness: Days 21 and 31\label{sect_pre_peak}}

 To highlight the spectroscopic peculiarities of SN 1994W,  we first review the spectral evolution of Type II-P SNe
 which are governed by ejecta cooling and  the increased effects of metal line blanketing.
The SED is blue during the first 10 days after shock breakout
and reddens dramatically as hydrogen recombines in the ejecta. Balmer lines are
the strongest lines at all times, initially accompanied by He{\sc i} lines
(5875\,\AA\ is the most conspicuous, but other optical He{\sc i} lines are present)
and then by Fe{\sc ii} lines. These proceed in a monotonic sequence, so that He{\sc i} and Fe{\sc ii}
lines are not seen simultaneously.

 In Type IIn SNe, and by contrast with Type II-P SNe, such smooth and monotonic evolution does not systematically hold.
The spectrum of SN 1994W on day 21 shows the simultaneous presence of
both Fe{\sc ii} and He{\sc i} optical lines (with absorption and emission components),
in combination with a blue continuum.
To illustrate this peculiarity,  we show in Fig.~\ref{fig_94W_0804} three different fits
to the observations of SN 1994W on day 21.
In the left panel, we use a ``hot model'' with the following properties:
$R_{\rm phot}=2.86 \times 10^{15}$\,cm, $V_{\rm phot}=$830\,\kms, $T_{\rm phot}=$10350\,K,
$\rho_{\rm phot} = 8.2 \times 10^{-15}$\,g\,cm$^{-3}$,
$N_{\rm e,phot} = 3.3 \times 10^9$\,cm$^{-3}$, and $L_{{\rm OBS},R_{\rm Max}}= 6.17 \times 10^9$\,\lsun.
(We have scaled the synthetic flux by a factor of 1.06.)
The model ejecta are relatively hot, nearly fully ionized (only helium is partially ionized
just above the photosphere). The observed shape of the SED
is approximately matched, but there are severe discrepancies. He{\sc i}\,5875\,\AA\
is too strong, both in the emission strength and width (note that using
a solar composition for hydrogen and helium reduces, but does not resolve, this discrepancy).
Numerous lines around 4500\,\AA\ (Ti{\sc ii} and Fe{\sc ii}),
as well as around 5200\,\AA\ (mostly Fe{\sc ii}), are strongly underestimated.
Balmer lines are underestimated in strength, but well fitted in width.
The absorption at $\sim$-800\,\kms\ from line center is also well reproduced.

   In the middle panel of Fig.~\ref{fig_94W_0804}, we show the fits to the same observations,
but using a cooler model ($T_{\rm phot}=$7480\,K; we have scaled the synthetic flux by a
factor of 0.93), the one that fits the observations of SN 1994W
on 1994 September 1 (see Table~1 for characteristics). The slope
of the synthetic SED is now in greater disagreement, but the ionization seems more adequate,
as we predict all the observed lines, merely shifted vertically due to a redder/weaker continuum.
As an experimentation, we reduced the reddening from 0.17 to 0.05, and, as evident in the right
panel of Fig.~\ref{fig_94W_0804}, the fits to both the continuum and the lines become excellent.
We do not support this reddening, as it is incompatible with past works (SCL, C04) and with the
modeling done for later dates, even within the uncertainties. But, this experimentation suggests
that two distinct regions contribute, one to form a blue nearly-featureless SED (as we do not see
high-ionization lines), and another to produce the lines from low-ionization species (Fe{\sc ii}).
This is a distinctive feature of SN 1994W, not seen in Type II-P SNe, and only visible
on this day and on day 31.

  On day 31, the optical spectroscopic observations of SN 1994W undergo a drastic and atypical
change, with the SED becoming bluer (C04 stress that this change is genuine, and not the result
of a poor relative flux calibration), the narrow lines of Fe{\sc ii} appearing weaker
at, e.g., 5018\,\AA\ and 5169\,\AA. We identify H{\sc i} Balmer lines, He{\sc i}\,5875\,\AA,
He{\sc i}\,6678\,\AA, and He{\sc i}\,7065\,\AA, and Mg{\sc ii}\,7877--7896\,\AA.
In the cool model (same as that used to model observations on the 1994 September 1)
shown in the bottom panel of Fig.~\ref{fig_94W_0814}, the Mg{\sc ii} lines and
He{\sc i}\,5875\,\AA\ are well reproduced, but He{\sc i}\,7065\,\AA\ is not. By contrast, in the hot model
(same as the hot model used on day 21 presented above) shown in the top panel, the reverse is true.

  Overall, our fits to these pre-peak brightness observations are very poor.
The apparent increase in ionization between days 21 and 31 is a unique feature
of SN 1994W, never quite seen in standard SN ejecta (of any type), which are governed
by the cooling associated with expansion, typically mitigated by the release of stored internal
energy. The non-monotonic SED evolution of SN 1994W between days 21 and 31,
and the simultaneous presence of lines suggestive of both
high and low ejecta temperatures point towards two distinct radiating regions (or material),
of distinct properties.

 As there are clear inconsistencies between our model and the observations on days 21 and 31,
the quoted model parameters  do not accurately describe SN 1994W on these two dates.
It is possible, for example, that the photospheric radius and temperature derived for day 49
also apply at the earlier dates; the extra luminosity would then arise from an optically thin,
and more highly ionized, outer region.

%
%

  \subsection{Observations post-peak optical brightness: days 49, 57, 79, and 89.\label{sect_post_peak}}

\begin{figure}
\epsfig{file=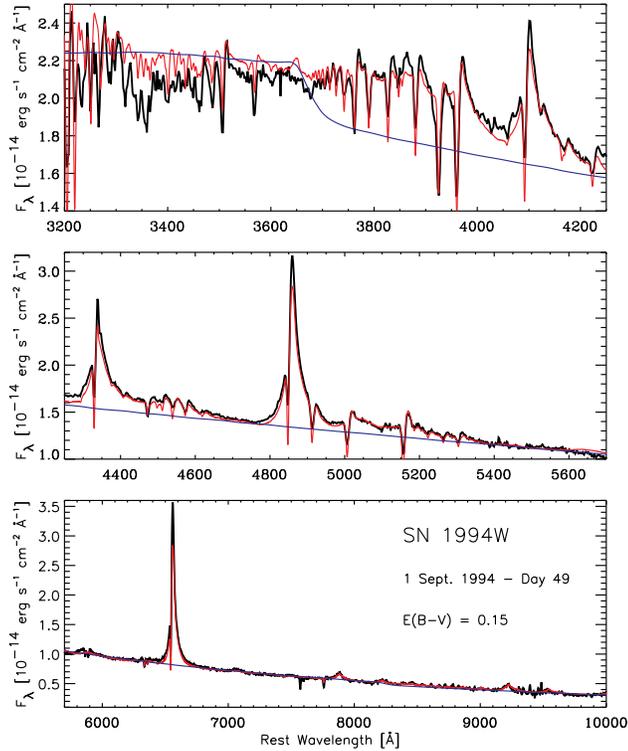,width=9cm}
\caption{Comparison between the reddened (E(B-V)=0.15\,mag) full (red) and continuum-only (blue)
synthetic spectra and the observations of SN 1994W on 1994 September 1 (day 49; black).
The synthetic flux is scaled by a factor of 0.87 to adjust to the absolute level
of the observed flux.
}
\label{fig_94W_0901}
\end{figure}

\begin{figure}
\epsfig{file=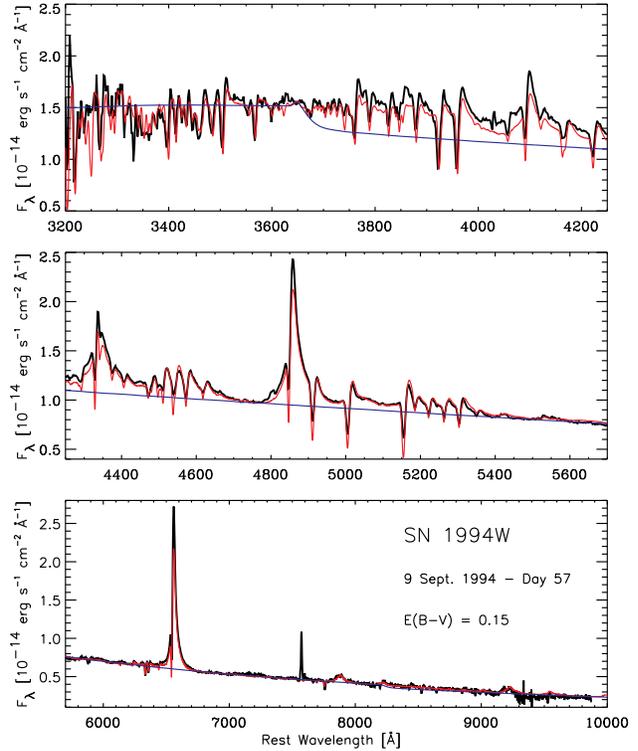,width=9cm}
\caption{Comparison between the reddened (E(B-V)=0.15\,mag; this is to reduce a slight discrepancy around
4000-5000\,\AA) full (red) and continuum-only (blue)
synthetic spectra and the observations of SN 1994W on 1994 September 9 (day 57; black).
The synthetic flux is scaled by a factor of 1.02 to adjust to the absolute level
of the observed flux.
}
\label{fig_94W_0909}
\end{figure}

\begin{figure}
\epsfig{file=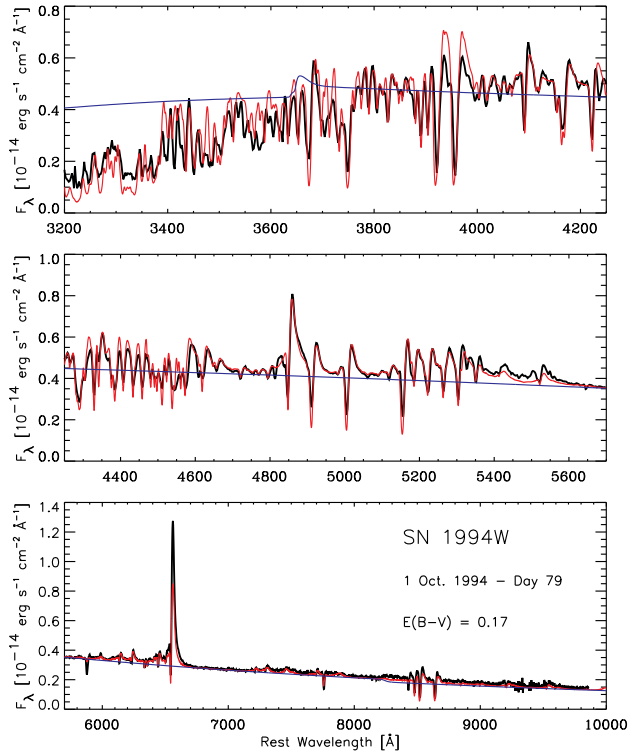,width=9cm}
\caption{Comparison between the reddened (E(B-V)=0.17\,mag) full (red) and continuum-only (blue)
synthetic spectra and the observations of SN 1994W on 1994 October 1 (day 79; black).
}
\label{fig_94W_1001}
\end{figure}

\begin{figure}
\epsfig{file=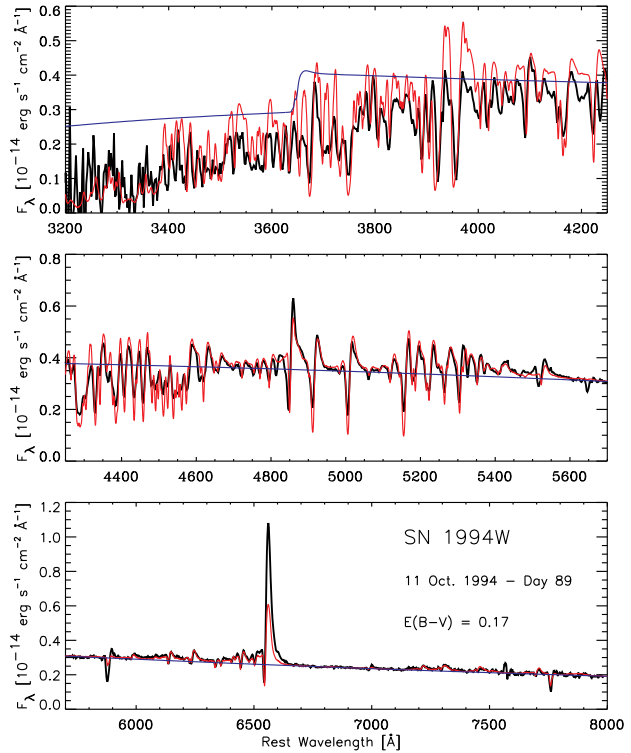,width=9cm}
\caption{Comparison between the reddened (E(B-V)=0.17\,mag) full (red) and continuum-only (blue)
synthetic spectra and the observations of SN 1994W on 1994 October 11 (day 89; black).
The synthetic flux is scaled by a factor of 0.87 to adjust to the absolute level
of the observed flux.
}
\label{fig_94W_1011}
\end{figure}

  From day 49 onwards, the visual brightness of SN 1994W decreases slowly, from $\sim$13.5
on that day down to $\sim$15 on day 110 (SCL). Over that time, the observed spectral evolution
is smooth and slow, analogous to
that of Type II-P SNe during the plateau phase, and our fits are satisfactory.
We present synthetic fits to observations in Fig.~\ref{fig_94W_0901} for day 49,
in Fig.~\ref{fig_94W_0909} for day 57, in Fig.~\ref{fig_94W_1001} for day 79,
and in Fig.~\ref{fig_94W_1011} for day 89. We also give the corresponding
model parameters in Table~1.
These spectra are dominated by lines of low-ionization species, such as H{\sc i}, Fe{\sc ii},
Na{\sc i}, Ca{\sc ii}, Mg{\sc ii}, with He{\sc i}\,5875\,\AA\ disappearing after day 59
(see tables provided as online material for a complete census of all contributing lines).
Observations and our synthetic SED now agree well, suggesting both lines and continuum form
in the same region of space, i.e., under similar conditions of density and temperature.
As on days 21 and 31, Balmer lines continue to show conspicuous
narrow line cores with extended wings. Our fits to the line cores tend to be underestimated
(by up to a factor of two on day 89), but we can reproduce well the strength and width of the profile wings.
These profile wings stem from multiple electron-scattering events of photons originally
trapped in the line core, a phenomenon, in our approach, which occurs in the photospheric
region (see \S\ref{sect_es}). We also predict the absorption dip at $\sim$-700\,\kms, coincident
for all lines within $\pm$100\,\kms. This latter feature is also associated here with absorption
internal to the photospheric region.  We go back to these two intriguing characteristics
in \S\ref{sect_es}. The underestimate of the narrow line H$\alpha$ flux
could be due to the neglect of time dependent effects (Dessart \& Hillier 2008),
or might be due to a change in the density distribution within the SN envelope
(as characterized by the exponent $n$).

\begin{figure}
\epsfig{file=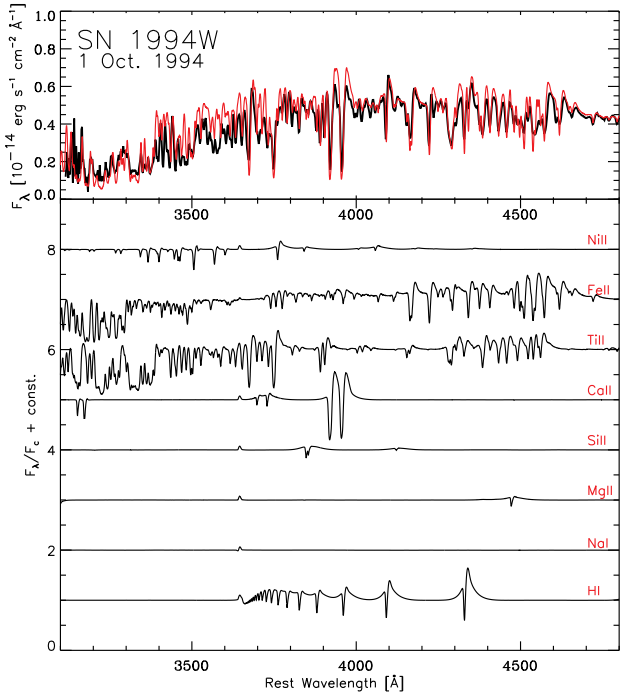,width=9cm}
\caption{
{\it Top:} Comparison between the reddened (E(B-V)=0.17\,mag) synthetic spectrum (red) and
the observations of SN 1994W on 1994 October 1 (day 79; black), between 3100 and 4800\,\AA.
{\it Bottom:} Rectified synthetic spectra for the model shown at top, but including bound-bound
transitions only of the individual species labeled on the right.
}
\label{fig_94W_line1}
\end{figure}

\begin{figure}
\epsfig{file=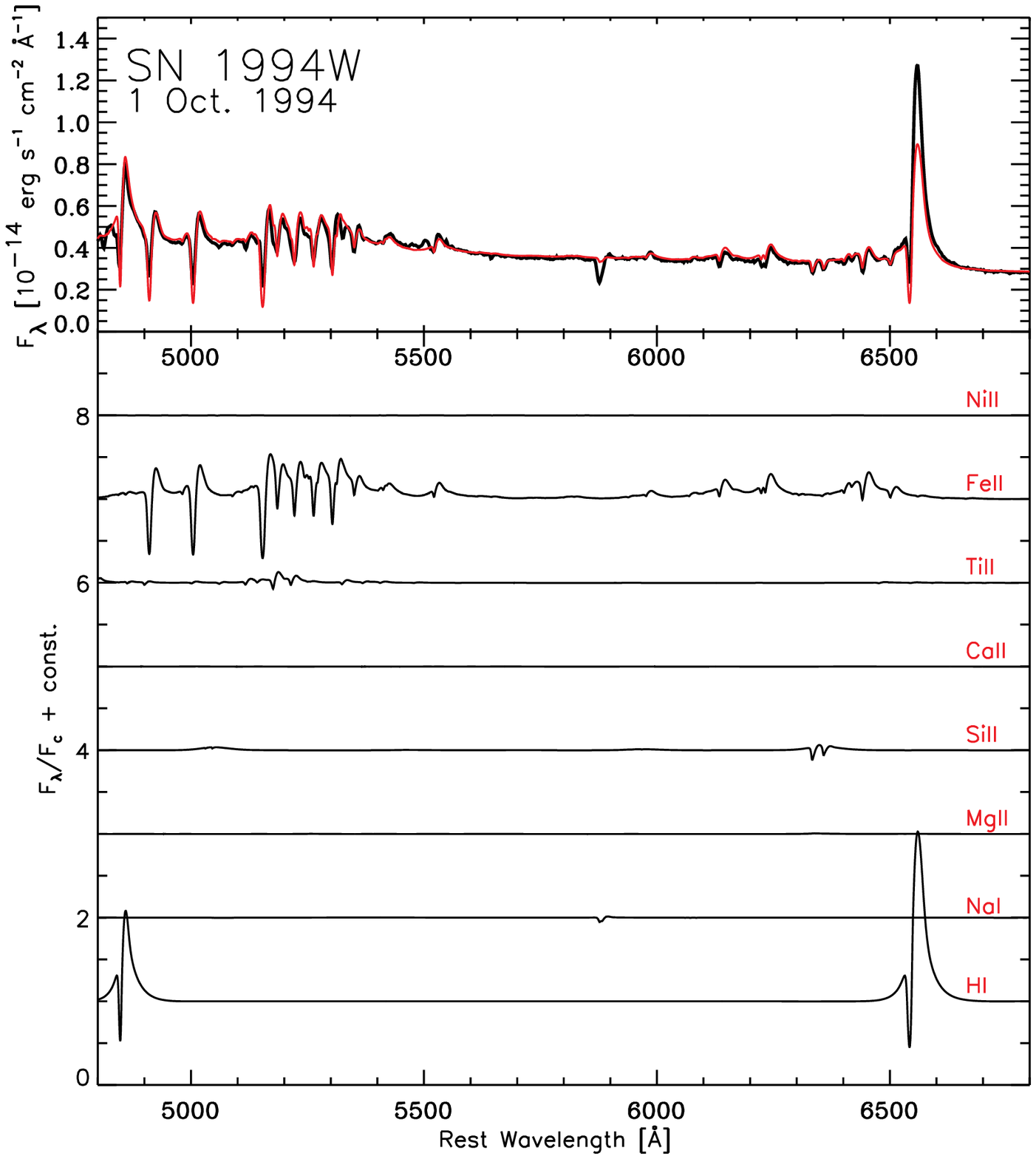,width=9cm}
\caption{
Same as Fig.~\ref{fig_94W_line1}, but for the range 4800 to 6800\,\AA.
\label{fig_94W_line2}
}
\end{figure}

\begin{figure}
\epsfig{file=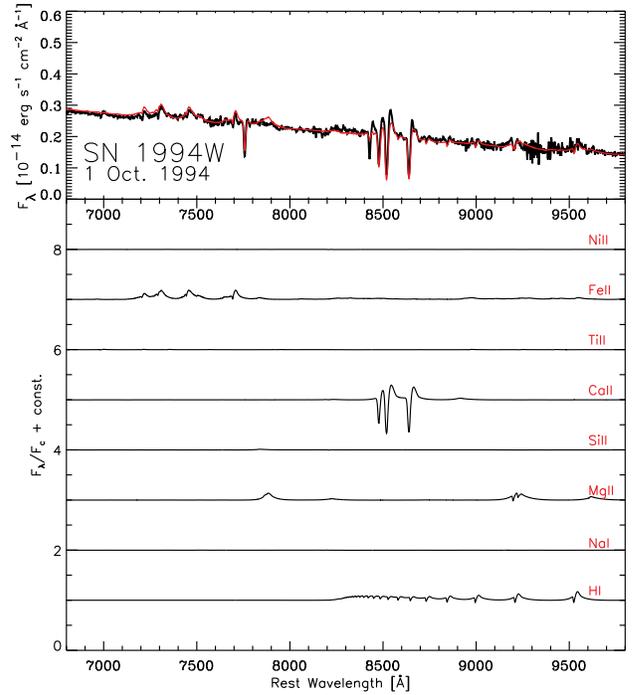,width=9cm}
\caption{
Same as Fig.~\ref{fig_94W_line1}, but for the range 6800 to 9800\,\AA.
\label{fig_94W_line3}
}
\end{figure}

In the top panel of Figs.~\ref{fig_94W_line1}--\ref{fig_94W_line3}, we reproduce the fit shown
in Fig.~\ref{fig_94W_1001} but this time zooming in on the 3100--4800\,\AA\ region
(Fig.~\ref{fig_94W_line1}), the 4800 to 6800\,\AA\ region (Fig.~\ref{fig_94W_line2}), and the
6800 to 9800\,\AA\ region (Fig.~\ref{fig_94W_line3}). Moreover, in the bottom panels, we
present the contributions of individual species by plotting the rectified synthetic spectra
obtained by accounting for bound-bound transitions of individual species, labeled at right.
At such a late time, besides Balmer and a few isolated Mg{\sc ii}, Na{\sc i}, Si{\sc ii}, and
Ca{\sc ii} lines, we note the presence of a forest of Fe{\sc ii} and Ti{\sc ii} lines
blueward of $\sim$4500\,\AA. The strengthening contribution of line emission and
absorption is also evident in Figs.~\ref{fig_94W_0901}--\ref{fig_94W_1011}, where the
blue curve describing the continuum SED departs more and more with time from the
red curve including all absorption and emission processes, the more so at shorter wavelengths.
This makes the continuum level difficult to assess, and the comparison with a blackbody
increasingly inadequate. This complication should be kept in mind when using blackbody
arguments.

%
%

Over this period from day 49 to 89, which covers the visual fading of the SN by $\sim$1.3\,mag,
we find a modest reduction in the photospheric temperature, of $\sim$1400\,K ($\sles$20\%), but a
large reduction in the photospheric radius, by $\sim$2$\times$10$^{15}$\,cm (a factor of 2 reduction
compared to the value on day 49). In the observer's frame and over these 40 days,
this corresponds to a motion of the photosphere inward at an average velocity of $\sim$6000\,\kms.
This is analogous to the reduction in the photospheric radius at the end of the plateau phase of Type II-P
SNe, in which an ejecta extending out to a few times 10$^{15}$\,cm becomes entirely optically-thin in
just $\sim$2 weeks. Heuristically, the photometric fading and the strongly correlated $B$ and $V$ magnitudes
support a reduction in the photospheric radius and a fairly constant photospheric temperature, respectively.

The dynamics of the CDS is that of linear expansion with time (C04). Our findings suggest that
the photosphere does not track the CDS, otherwise this linear expansion in time
at near constant photospheric temperature would correspond to a phase of
brightening. We instead observe a fading of the SN over that period. C04 associate the broad wings with
the CDS deducing an expansion velocity of 4000\,\kms. Over 100 days, the CDS would expand by
$\sim$3.5$\times$10$^{15}$\,cm, compared to our inferred reduction in $R_{\rm phot}$ of
$\sim$2$\times$10$^{15}$\,cm .


Our models suggest that as time progresses, the photosphere recedes to deeper and deeper layers in a
cold shell --- probably the SN ejecta (or more generally the inner shell) but possibly also associated
with the CDS initially.  We surmise that, as time goes on, the emitting material radiates and expands
sufficiently to cause ejecta cooling and recombination. The photosphere is completely slaved to the
layer of ionized material above which free electrons are too scarce to provide any sizable optical
depth.
Lines, which form above the photosphere, tend also to track this region of high density
and high ionization, in particular those that form primarily through recombination, e.g. Balmer lines.
Since the ejecta are hydrogen-rich,
this occurs at a temperature of $\sim$7000\,K, and that temperature is essentially fixed.
With further cooling, it is not the temperature of the photosphere but instead the radius of the
photosphere that adjusts, shrinking to deeper layers.

This process has already been identified as the primary cause for the plateau phase of Type II-P SNe:
the photospheric temperature remains fixed at $\sim$7000\,K, with a recession in mass coordinate of the
photosphere, compensated in this case by the fast expansion of the ejecta so that the photospheric
radius remains roughly constant during that phase (see Fig.~16 in Dessart et al. 2008).
The end of the plateau phase and the fast drop into the nebular phase will correspond to the time
when the inward-traveling photosphere no longer sustains a high enough density and ionization to
remain optically thick. In other words, the end of the plateau phase would not correspond to the
time when the CDS reaches the outer edge of the outer shell (C04), but to the time when the
photosphere eventually reaches the cold, fully-recombined, inner layers of the SN ejecta/inner-shell.

Since the CDS seems to be optically thin after the epoch of peak-brightness, it cannot obscure
the SN ejecta buried at depth. The absence of broad SN-like spectral features may then result
from the ejecta deceleration  by the reverse shock. Some deceleration is expected if conversion
of kinetic energy into internal energy is to power the SN brightness, but the exact circumstances
that cause the absence of ejecta material with velocity at the photosphere greater than $\sim$800\,\kms
at all times are unclear. Even at day 121, in the nebular phase, the expansion velocities indicated
by H$\alpha$ are consistent with an expansion of $\sim$800\,\kms (C04). This is consistent with the idea
that we are observing a shell of material moving with near constant velocity, but which is now in
the nebular phase.



\section{Discussion of line profile formation and the electron scattering wings}
\label{sect_es}

\subsection{Observational aspects}

\begin{figure*}
\epsfig{file=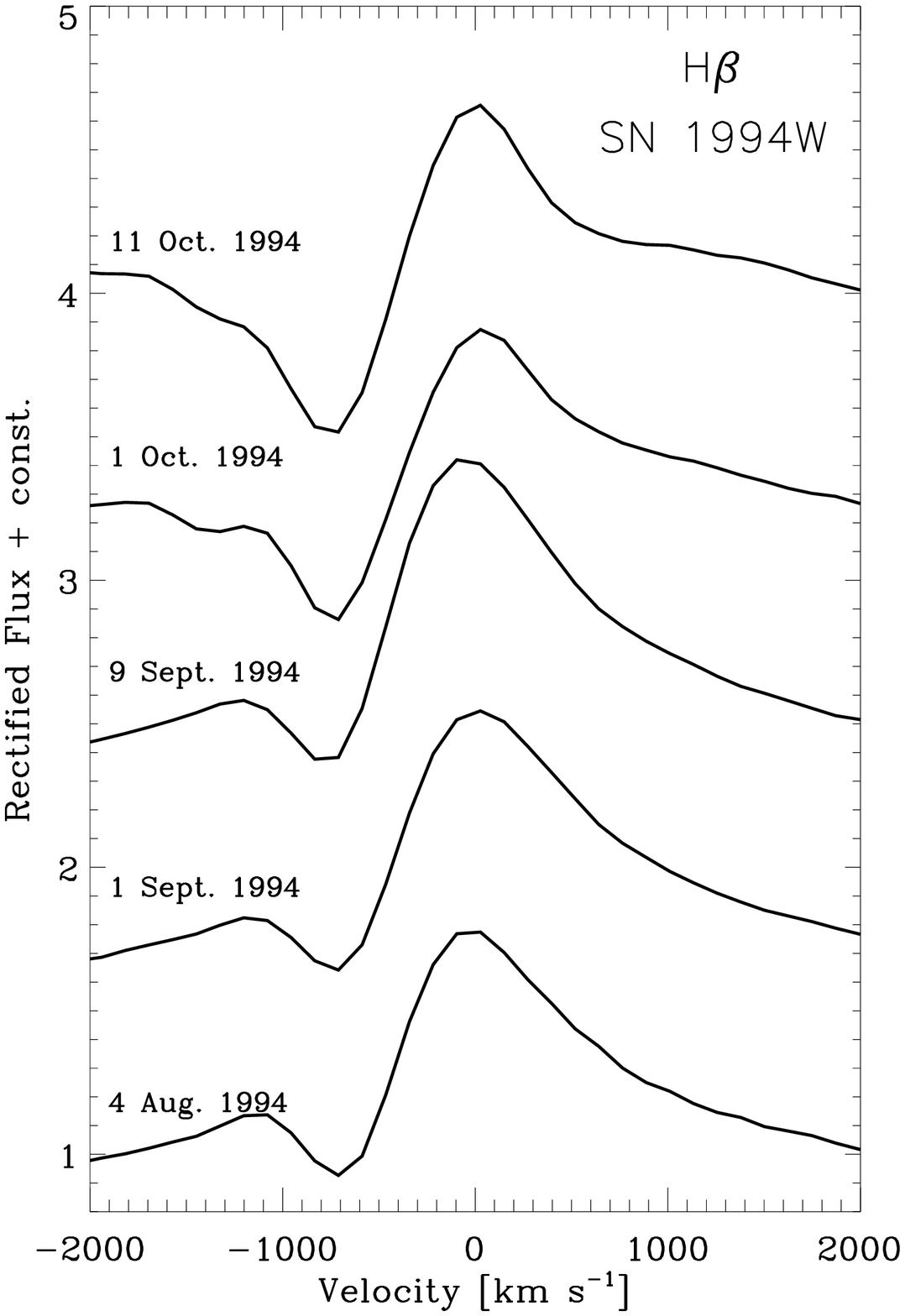,width=8cm}
\epsfig{file=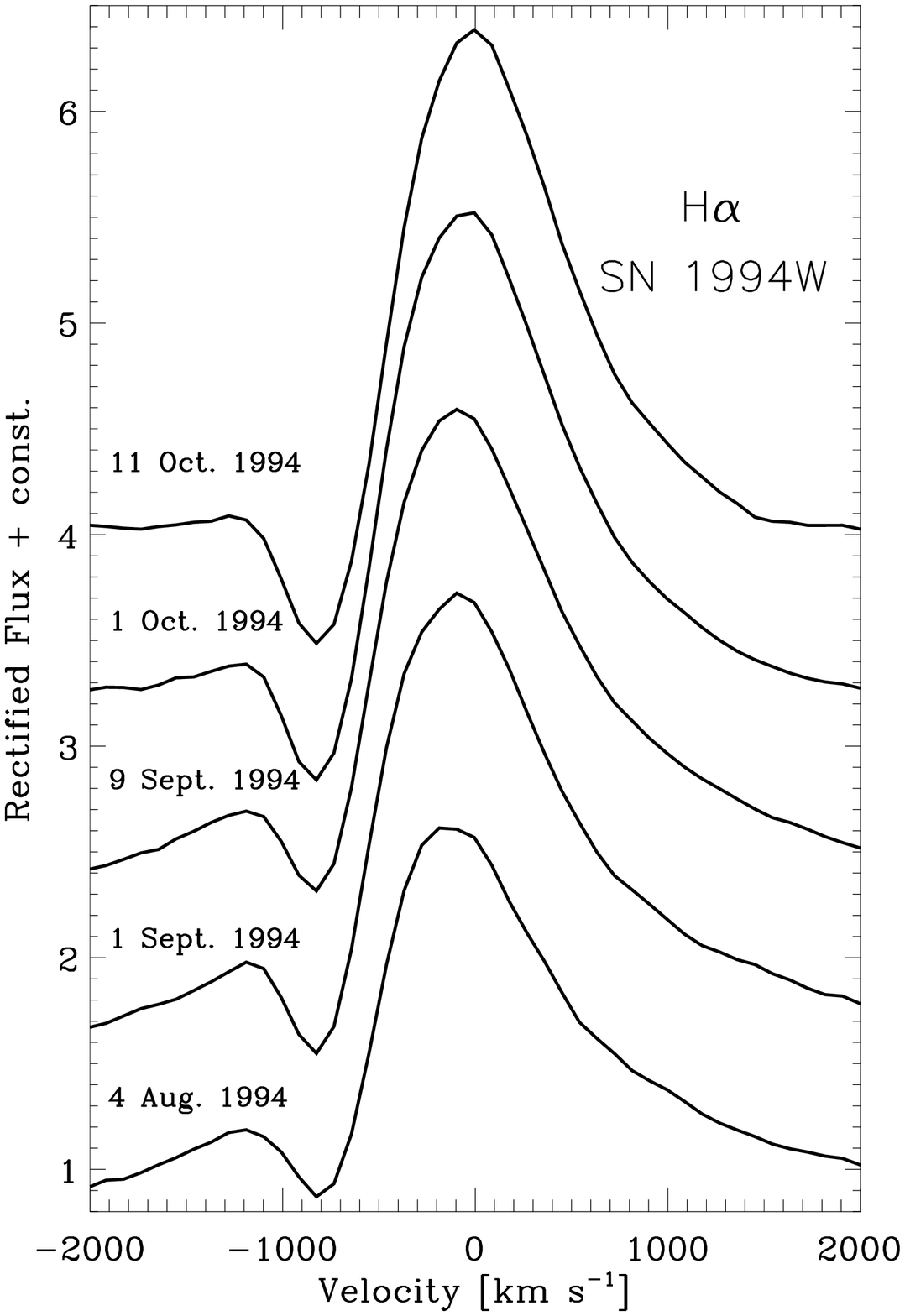,width=8cm}
\caption{
Montage of the evolution (a label gives the date) of the observed H$\beta$ ({\it left})
and H$\alpha$ ({\it right}) line profiles versus Doppler velocity (we correct for a redshift of
1249\,km\,s$^{-1}$)
for SN 1994W (the spectra are normalized to unity at +2000\,km\,s$^{-1}$ and shifted
vertically for visibility). Note the extended wings of the narrow-peak Balmer line profiles,
with a blue-wing emission contribution that decays with time, by contrast with
the strengthening of the red contribution.
Note also the velocity blueshift of the emission peak at early times (see Dessart \& Hillier 2005a
for a physical interpretation), as well as the near-constancy of the velocity location of maximum absorption.
\label{fig_94W_ha_hb}
}
\end{figure*}

\begin{figure}
\epsfig{file=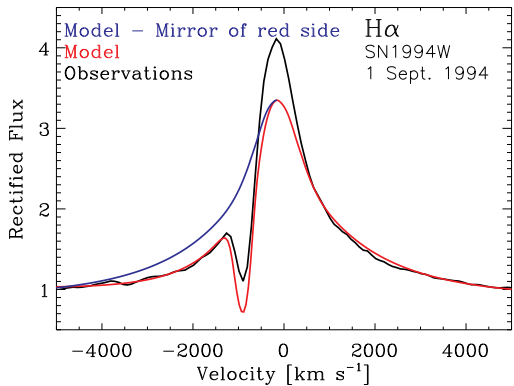,width=8cm}
\caption{
Comparison between the observed (black) and synthetic (red) H$\alpha$ line profile
for 1994 September 1 (see also Fig.~\ref{fig_94W_0901}).
Our model of a single formation region for the continuum and lines
reproduces the presence both of a narrow line core (somewhat underestimated)
and broad line wings. Numerous line profiles in the spectrum of SN 1994W share this morphology.
To illustrate the stronger red wing flux, we mirror the red part of the
H$\alpha$ profile and draw it in blue on the blue side.
\label{fig_CMFGEN_obs_Halpha}
}
\end{figure}

The distinctive feature of Type IIn SNe is the presence of narrow, and usually symmetric,
line cores at all times. In some cases, as for SN 1994W, these are accompanied by broad, and relatively
symmetric line wings, giving the  overall line profile a triangular shape (see Fig.~\ref{fig_94W_ha_hb}).
These are in stark contrast with the broad and P-Cygni profiles, observed in Type Ia/b/c and Type II-P SNe,
associated with optically-thick line formation in an
expanding medium. As shown in the preceding sections, our modeling approach allows us to
reproduce this hybrid morphology of a narrow line core and extended line wings, as illustrated more
clearly in our fit to the H$\alpha$ line profile on 1994 September 1 (Fig.~\ref{fig_CMFGEN_obs_Halpha}).

In Type IIn's, the narrow line core has been interpreted as line emission from a slowly expanding
region. C04 argue that the lack of SN features in the optical spectrum of SN 1994W is in support
of the presence of an {\it optically-thick} CDS, whose velocity is on the order of $\sim$4000\,\kms.
The narrow line core flux then stems from the region between the CDS and the outer slowly
moving circumstellar shell. The broad wings, according to C04, may stem from various mechanisms/regions.
Turbulent, clumpy, and fast moving material above the CDS and underneath the outer shell may contribute.
Radiation-driving may also cause some acceleration of the material just above the CDS, which could
then be the origin of some line broadening.

C04 also argue for multiple scattering of line photons by electrons in
an optically-thick CS envelope. Such non-coherent electron scatterings redistribute
line photons in frequency space, and if multiple, can cause a spreading of the line core flux (see below).
Because multiple-scattering of line photons occurs when the electron-scattering optical depth is large,
the presence of such wings has been associated with the presence of an external optically-thick CS envelope.
The decrease of the strength of these wings as time progresses (Fig.~\ref{fig_94W_ha_hb})  would then follow
from the decrease in the shell thickness above the outward-migrating CDS.\footnote{
Note that as the optically-thick radiating layer migrates outward,
the optical-thickness of that external shell above it should decrease and
eventually vanish. By contrast, line profiles always show narrow line cores
and broad line wings, over the 80-day period covering the bright phase of the
supernova.}

\begin{figure*}
\epsfig{file=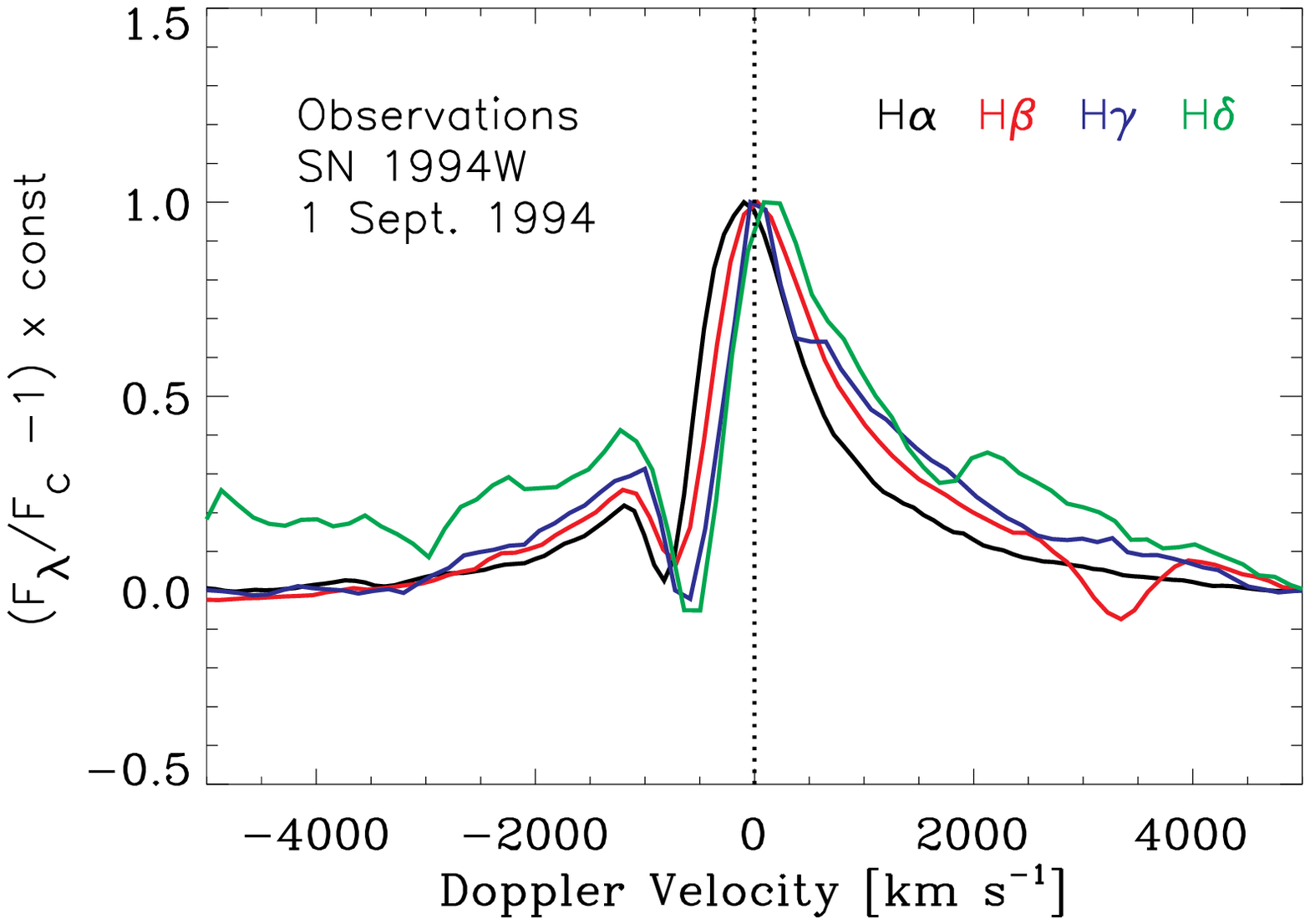,width=8cm}
\epsfig{file=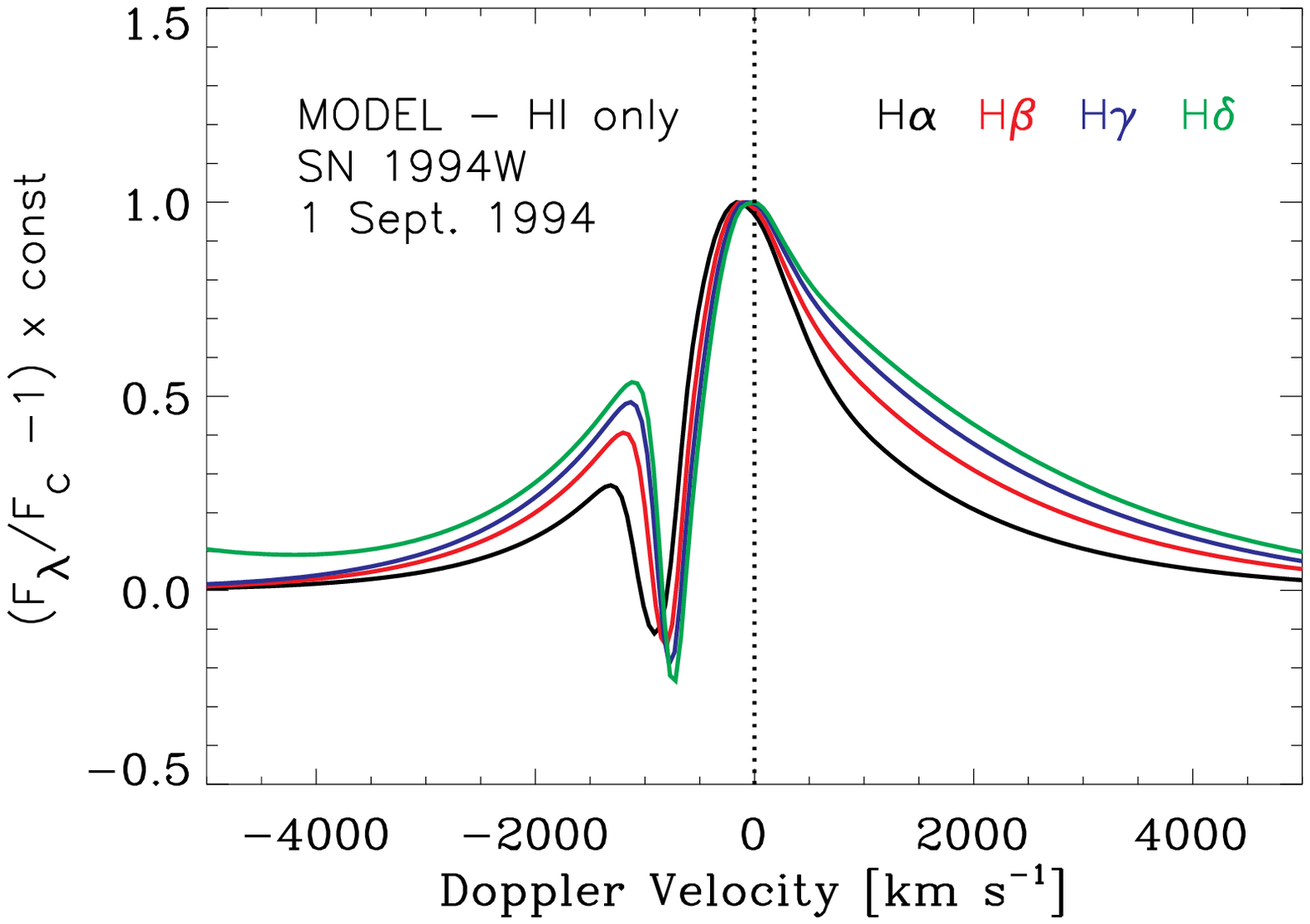,width=8cm}
\epsfig{file=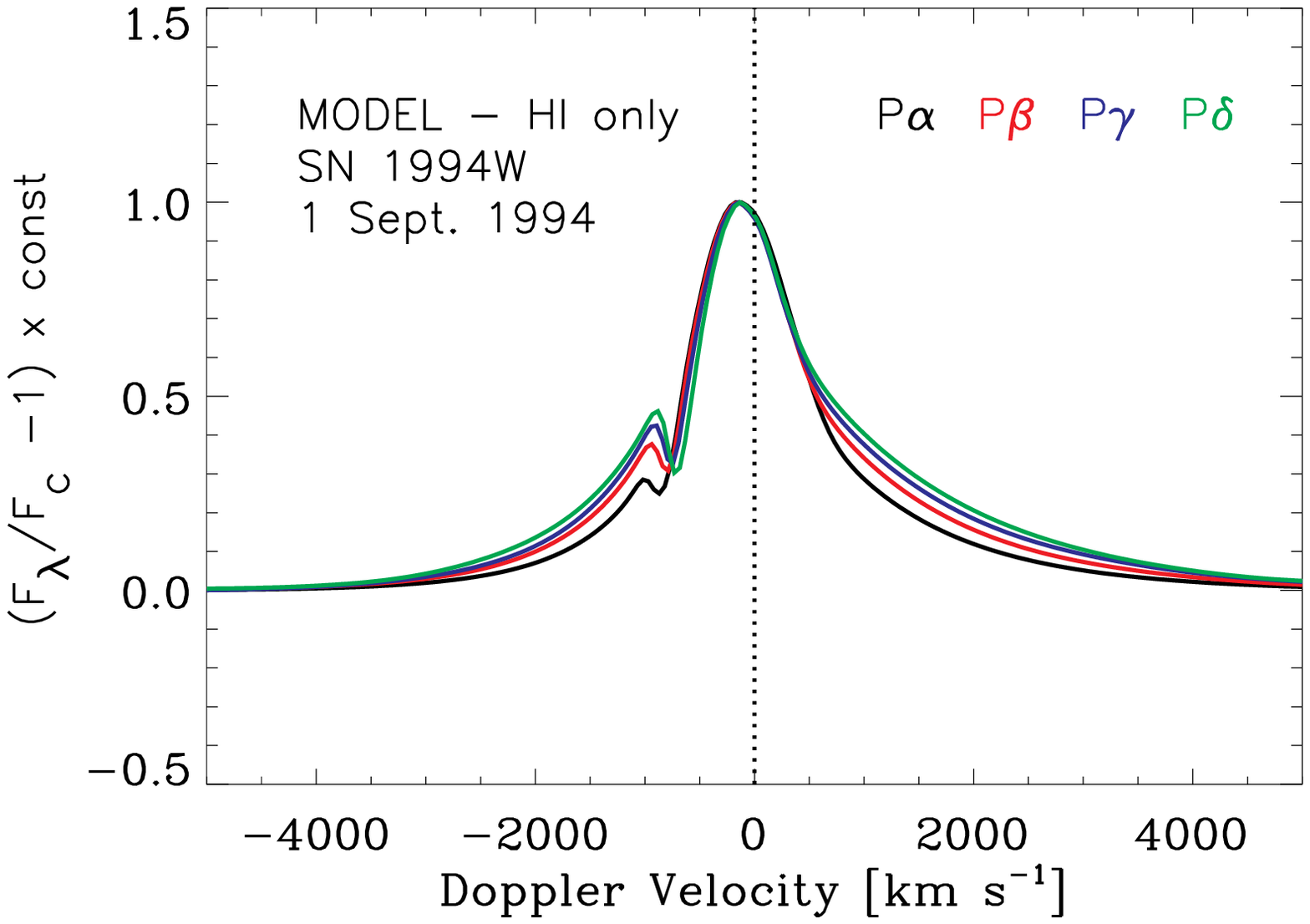,width=8cm}
\epsfig{file=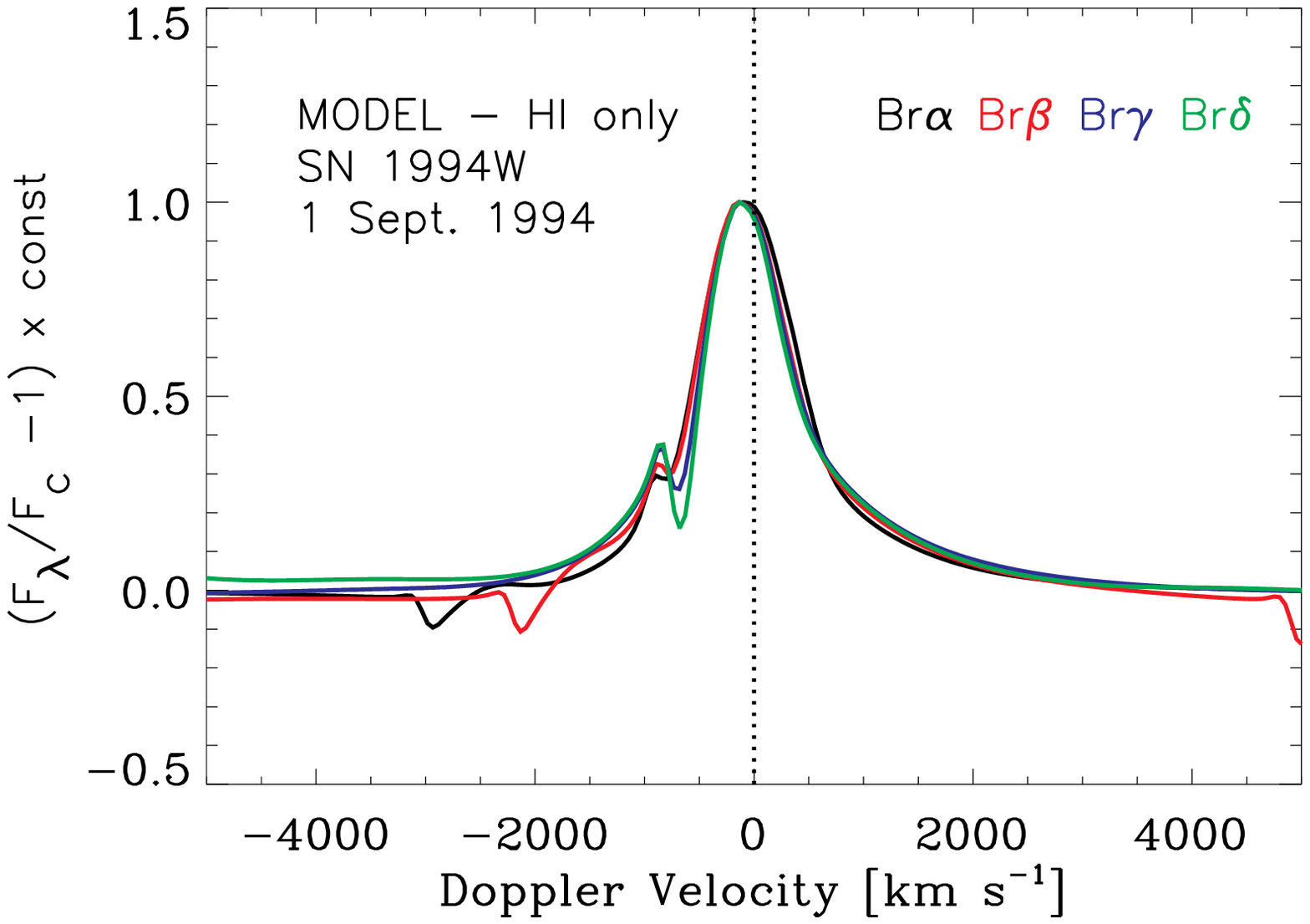,width=8cm}
\caption{{\it Top:} Observed ({\it left}) and theoretical ({\it right})
Balmer line profiles versus Doppler velocity
for the first four terms of the series ($\alpha$:black; $\beta$: red; $\gamma$: blue; $\delta$: green).
For a better comparison, we use the same ordinate range for all panels.
We also rectify the spectra and normalize to the peak flux for the corresponding line.
{\it Bottom:} Same as {\it top right}, but this time for the
Paschen ({\it bottom left}) and Brackett ({\it bottom right}) series.
For this illustration, we use the model for the observations of SN 1994W on 1994 September
1, as shown in Fig.~\ref{fig_94W_0901}, but include only the bound-bound transitions of
hydrogen (among the first 30 atomic levels).
Note how the flux in the line-profile wings gets relatively stronger compared to the peak value
as we move up the Balmer series, from H$\alpha$ to H$\delta$. We find that electron scattering wings
are stronger for lines that form at higher optical depth (see Fig.~\ref{fig_ep_tau_hi}).
If the profile wings were due to an {\it external} scattering layer, the flux in the profile wings
would instead scale linearly with the line flux.  Note that the blueshifted absorption at $\sim$-3000\,\kms
of Br$\alpha$ and at $\sim$-2000\,\kms of Br$\beta$ (right panel) are due to H{\sc i}\,4.0198\,$\mu$m and
H{\sc i}\,2.6119\,$\mu$m, respectively.\label{fig_CMFGEN_Balmer_model_obs}
}
\end{figure*}

Insights into the source of electron scattering can be provided by studying different H
lines. For example, a test for the presence of a shell external to the line formation region is that
the strength of the electron scattering wings should be directly proportional to the line
flux. Indeed, given the electron scattering optical-depth of the external shell, the number of line
photons that end up in the electron-scattering wings of each line should scale with the
number of line photons injected at the base of this optically-thick layer.
In the {\it top-left} panel of Fig.~\ref{fig_CMFGEN_Balmer_model_obs}, we show the
observed (rectified and scaled) Balmer line profiles on 1994 September 1.
Having normalized each peak line flux to unity, it is apparent that the flux in the profile wings
increases as we progress up the Balmer series, although for H$\delta$ the rectification may be somewhat
in error (there is a background of overlapping lines contributing in the H$\delta$ region).
This argues against an external shell as the sole source of electron scattering. However, the observed
behavior can be explained if the regions of line formation and electron scattering overlap.

\begin{figure}
\epsfig{file=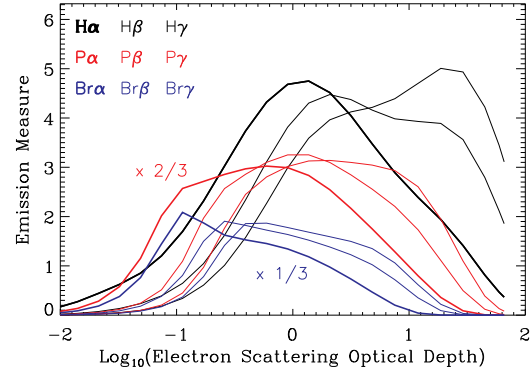,width=8cm}
\caption{Variation of the emission $\zeta$ (Hillier 1987) versus the Log of the electron-scattering optical
depth for Balmer (black; scaling of unity), Paschen (red; scaling of 2/3), and
Brackett (blue; scaling of 1/3) lines for the model shown in Fig.~\ref{fig_94W_0901}.
We draw higher series members with thinner lines.
The quantity $\zeta(R)$ relates to the total emission interior to $R$ in the line through the integral
$\int_{R_0}^{R} \zeta(R') d \log R'$. $\zeta(R_{\rm Max})$ is the total emission in the line.
Note how the site of emission resides deeper in, i.e., at higher optical depth, for higher
energy transitions in each series. There appears to be an increasing outward shift
from the Balmer to the Paschen and to the Brackett series.
In our synthetic spectra, electron scattering effects are stronger for lines that form deeper
in, as observed, and are internal to the photospheric region.\label{fig_ep_tau_hi}}
\end{figure}

In the {\it top-right} panel, we show the corresponding synthetic Balmer line profiles produced
with the models described in the previous section. These exhibit
the same correlation as the observed profiles --- the wings extend $\sim$3000\,\kms away
from line center, even for the weaker lines in the series, and are strongest  for the higher series members.
Hence, we do not identify a clear (linear) flux correlation between line core and line wing.
By contrast, we find that the synthetic line wing strength scales with the electron-scattering
optical depth in the formation region of the corresponding line. In Fig.~\ref{fig_ep_tau_hi},
we plot for Balmer lines (black; the thickness of the line distinguishes the transition)
the variation of the quantity $\zeta(R)$ (which corresponds to the emission interior to $R$
in the line through the integral $\int_{R_0}^{R} \zeta(R') d \log R'$)
with respect to the electron-scattering optical depth
integrated inwards from the outer grid radius.
The relatively weak (strong) line wings of H$\alpha$ (H$\beta$)
correlate with the relatively low (high) electron-scattering
optical depth in the formation region of the line. In this plot, the situation is most
severe for H$\gamma$, which forms deepest. Hence, despite the large H$\alpha$ flux, the wing flux in both
the blue and the red is relatively weaker  than in the other lines.

For completeness, we have also included in the bottom panels the predictions for the Paschen and
Brackett series. The enhanced scattered flux for high series members is also seen in the Paschen series,
but is not so obvious in the Brackett series. This is a key result --- Br$\alpha$ has a higher
optical depth and forms further out in the envelope than does H$\alpha$, and thus exhibits much
less obvious electron scattering wings. Observations of Paschen and Brackett lines would help provide
key information as to the nature of the broad wings.  If the layer is external the fraction of flux
in the line wings relative to the line core will be similar for the Balmer and Brackett series,
whereas if the electron scattering layer overlaps the line formation region the fraction of line
flux in the line wings will be greatest for the Balmer series. In addition, if a significant fraction
of the broad wings arise from intrinsic emission by a fast moving shell/envelope, one would anticipate
that the broad wings should be easily discernible in Br$\alpha$.


To conclude, we find that the strength of the Balmer line profile wings correlates
with the electron-scattering optical depth in the corresponding line formation region,
the lines forming at greater depths having relatively stronger flux in their wings.
In our approach, the broad line wings are thus formed within the photosphere rather than caused by
an external optically-thick shell.\footnote{Another effect could have arisen from
the alternate decay channels into transitions that have a lower optical depth.
For H$\alpha$, there is no alternate decay route for the upper state but to go into $n=2$.
An H$\beta$ photon can yield a couple of P$\alpha$-H$\alpha$ photons. For H$\gamma$, there are more
options with P$\beta$-H$\alpha$, Br$\alpha$-H$\beta$, or Br$\alpha$-P$\alpha$-H$\alpha$.
The potential escape allowed through non-coherent electron-scattering for H$\alpha$ could thus be more
important compared to higher transitions that have alternate decay routes. If this effect prevailed,
the flux in the electron-scattering wings of H$\alpha$ would be stronger relative to higher transitions
in the series, and this is not supported by observations.} In addition, and unlike C04,
we find no need for an extra emission source to contribute to the strength of the broad wings.

\subsection{Electron scattering theory}

In expanding media, electron-scattering introduces a systematic redshift of line photons
in the observer's frame. This is particularly evident in strong spectral lines observed in
SNe and hot star spectra. However, to the steep density fall-off in SN ejecta corresponds
a steep drop-off in the electron-scattering optical depth, so that generally only a little
velocity contrast exists between the emission site of a line photon and the location where
it is scattered by a free-electron. Free streaming of such line photons prevails soon above
the photosphere. The extent of the red-wing electron-scattering in SN spectral lines is
therefore modest, at least in comparison with what is seen in hot star winds and their
characteristic $1/R^2$ density distribution.

A second effect, present even in a medium that is globally at rest, is caused by the relatively
large thermal velocity ($V_{\rm th,e}= 550 \sqrt{T/10^4}$\,\kms) of free electrons, even
at moderate temperatures.
%
%
Scattering with free electrons will lead to appreciable Doppler shifts of line photons relative
to the narrow intrinsic width of the line which is typically a few \kms (for an ion with atomic
weight $A$,  $V_{\rm Dop}=12 (T/A10^4)$\,\kms).
Upon scattering out of the line core, the photon may be redistributed in frequency into the line wing
where the line optical depth is lower, and thus escape entirely from the line. It is then subject
to the electron-scattering opacity which prevents it from free-streaming to infinity; the photon
may experience multiple scatterings with free electrons (with blueshifts/redshifts) before escaping.
The frequency shifts may add or cancel but the cumulative statistical effect will be to lead
to an appreciable flux at many electron Doppler widths from line center, the more so for larger
electron-scattering optical depth at the emission site of the photon.

In fast moving SN ejecta, thermal motions are small relative to the expansion velocity; thus the influence
on the observed line profiles of incoherent scattering due to the electron thermal motions is small.
However, in slower moving SN ejecta the frequency shifts due to the thermal motions of the electrons
dominate over that due to expansion, and strong red and blue wings are observed. Lines forming deeper
in the photosphere, at higher electron-scattering optical depth, show the strongest wing to peak flux
ratio, as observed. Here, these numerous electron-scattering events occur internally to the photosphere,
rather than in an outer optically-thick shell.


In our SN models for SN 1994W, two important effects were observed --- the profile shape, and
to a lesser extent the equivalent width (EW), were strongly dependent on the number of electron scattering iterations
that were performed. To understand this finding we first need to explain the computational procedure
used to compute the line profiles.

To compute the line profiles we first solve for the atmospheric structure and level populations.
These are then used by a separate program, CMF\_FLUX (Busche \& Hillier 2005), to compute the
observed spectrum. For the first iteration, coherent scattering in the comoving frame is assumed.
This assumption ``conserves'' scattered line photons and allows for redistribution effects due
to the expansion of the SN envelope. Using the newly computed mean intensity J we now allow for
the effects of frequency redistribution by electron scattering using the technique of
Rybicki \& Hummer (1994), and recompute the mean intensity
in the comoving frame. To allow for the effects multiple scattering has on the frequency redistribution
of line photons, it is necessary to iterate. For fast moving SNe, two iterations is generally sufficient,
but for the slow moving SNe, we sometimes had to perform 10 to 20 iterations to get converged line profiles.
For accurate calculations the same Doppler width should be used for both the line-source function and
the line profile calculations. However, for normal SNe changes in the adopted Doppler width have only
a minor (and well understood) effect on the line profile.


To further assist in understanding the observed behavior we also need to consider the relevant scales
for both line formation and electron scattering. Since we have a power law density distribution,
with exponent $n$, the characteristic length scale for electron scattering is $\sim R/(n-1)$.
A characteristic scale over which a photon interacts with a line in an expanding medium is the
Sobolev length (Sobolev 1960) defined by
  \begin{equation}
    L_{\rm SOB} = V_{\rm th, eff}/|dV/dR|,
  \end{equation}
where
\begin{equation}
V_{\rm th, eff} = \sqrt{ V_{\rm th,i}^2 +  V_{\rm turb}^2}
\end{equation}
is the effective ion thermal velocity, $V_{\rm th,i}$ is the ion thermal velocity, and $V_{\rm turb}$
is a microturbulent velocity that accounts for small-scale turbulent motions. The Sobolev length gives
the radial scale over which the velocity changes by $V_{\rm th, eff}$. In other words, it is the approximate
size of the resonance zone in which a photon is trapped in a line. In our simulations, we typically
adopt $V_{\rm turb}=$50\,\kms, which spreads the line over a broader frequency range than that given by
its intrinsic width of a few \kms.

In the SN context, we have $L_{\rm SOB}/R_{\rm phot} \approx V_{\rm th, eff}/\beta V_{\rm phot}$.
The Sobolev length may thus vary through changes in the velocity gradient (controlled by the parameter
$\beta$) or in the ejecta velocity (controlled by $V_{\rm phot}$)\footnote{The Sobolev length, in general,
varies with direction of photon travel, and for photons traveling perpendicular to the radius vector
it is always $R V_{\rm th, eff} / V$, independent of the velocity gradient. For simplicity
we retain the formulation above but stress that the average Sobolev length has a much weaker dependence
on $\beta$ than does the radial Sobolev length.}. In the Type II-P SN 1999em after a few weeks,
$V_{\rm phot}$ is on the order of $\sim$4000\,\kms, compared to $\sim$800\,\kms in SN 1994W.
Normalized to $R_{\rm phot}$ and assuming homologous  expansion ($\beta=$1), $L_{\rm SOB}$ is five times
larger in SN 1994W compared to SN 1999em.


For SN1994W, $L_{\rm SOB}/R_{\rm phot}$=50/800$=$1/16 which is only slightly smaller than the
electron scattering scale length.  As this path length is comparable to the electron scattering scale height,
there is a significant probability that a photon will be scattered by an electron within the resonance zone.
Further, this can facilitate the escape of line photons as the scattering can effectively ``shift'' the photon
out of the resonance zone by altering its frequency. As we increase $L_{\rm SOB}$,  and assuming a fixed line
source function, more photons will be incoherently scattered by electrons in the resonance zone. As most of
these photons would normally be destroyed, more photons will escape and the line EW will increase.
Moreover, since these photons can be subsequently scattered elsewhere, it is necessary to run many iterations
to follow their redistribution in frequency space. Because the line EW depends on the adopted Doppler width
(and hence Sobolev length) the line source function should be computed using the same Doppler
width.\footnote{All synthetic spectra shown here are computed with a value of 10\,\kms for the turbulent velocity.
Note that to limit the computation time, this choice only applies when computing the emergent spectrum
from a formal solution of the transfer equation. When solving for the full radiation transport problem
including level populations, we assume a turbulent velocity of 50\,\kms, and generally assume that
electron-scattering is coherent in the comoving-frame (i.e., we ignore redistribution effects due to
the thermal motions of the electrons), so as to allow the linearization of the electron-scattering source
function (see Hillier \& Miller 1998 for details).
Tests we performed show that there is a weak sensitivity of the level populations and
the emergent radiation field to variations in the turbulent velocity from 10 to 50\,\kms -- the choice
of 50\,\kms minimizes computational effort and is numerically advantageous.}

A critical characteristic of Type IIn SNe compared with more normal Type II SNe is their small photospheric
velocity, which leads to both a larger Sobolev length, and a larger ratio of  $V_{\rm th,eff}/V_{\rm phot}$.
Both effects facilitate the appearance of broad wings. The larger Sobolev length facilitates the escape
of line photons by electron scattering,
even from locations where line photons might not normally escape, and potentially at
depths where the electron scattering optical depth is still significant. The larger ratio
$V_{\rm th,eff}/V_{\rm phot}$ makes the incoherent wings due to the thermal motions of the
electrons more apparent relative to the observed line width as set by the SN expansion velocity.

We now illustrate these two effects. In Fig.~\ref{fig_comp_vcore}, we show a comparison between
synthetic H$\alpha$ line profiles obtained with the model used to fit observations on 1994 October 11,
but differing in the adopted values of the base velocity $V_0$ (the ratio $V_{\rm phot}/V_0$ is the same
in all models; $\beta=1$ in all cases). Increasing $V_0$ yields broader line cores, the red wing becomes
stronger than the blue wing, and the profile looks increasingly asymmetric.  Hence, by merely decreasing
$V_0$ from 1500\,\kms to 150\,\kms,  we go from a broad P-Cygni profile typical for a Type II-P SN to a
symmetric line profile with a narrow line core and broad wings typical for a Type IIn SN.

In Fig.~\ref{fig_comp_flux_beta}, we show models with different
velocity distributions but varying the exponent $\beta$ entering the velocity law.
Recall that reducing $\beta$ first reduces the maximum velocity, but it also increases the Sobolev
length. Hence, by reducing $\beta$ one can reduce the redistribution of line photons to the red associated
with expansion, and increase the importance of electron scattering as a means of escape for line photons.
All three models shown in Fig.~\ref{fig_comp_flux_beta}
yield a quasi-symmetric profile that extends from line center out to $\pm$3000\,\kms, although the
maximum expansion velocity in the ejecta varies from 4000, to 1364, and 1123\,\kms, as we
reduce $\beta$ from 1, to 0.2, and 0.01, respectively.
Non-coherent electron scattering is thus the primary line
broadening mechanism, so that, paradoxically, the broad line wings testify for slow rather than fast
expansion. For smaller $\beta$, the blue-wing emission strengthens and arises exclusively from
redistribution to the blue of photons in the line core. This occurs at the expense of the line core flux,
which indeed decreases correspondingly.

%
%
\begin{figure}
\epsfig{file=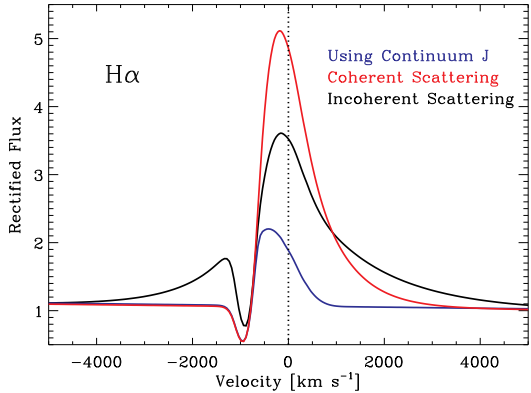,width=8cm}
\caption{
Comparison between the synthetic H$\alpha$ line profile
obtained assuming incoherent (black) or coherent (red) electron scattering in the comoving-frame.
In blue, we show the effect of using the continuum (rather than the line)
mean intensity when computing the electron-scattering source function, which shows
that a large fraction of line photons suffer at least one scattering
with free electrons as they escape from their emitting region.
For this illustration, we use the same model that fits the observations of SN 1994W on
1994 September 1, whose characteristics are given in Table~1.
\label{fig_CMFGEN_Halpha_es}
}
\end{figure}


Another element of interest in the observed Balmer line profiles shown in the top left panel
of Fig.~\ref{fig_CMFGEN_Balmer_model_obs} is the location of the P~Cygni profile dip, which ranges
from $-900$\,\kms\ for H$\alpha$ to $-500$\,\kms\ for H$\delta$ on 1994 September 1.
The velocity shift of this absorption component is well reproduced by our synthetic line profiles,
as shown in the top-right panel of Fig.~\ref{fig_CMFGEN_Balmer_model_obs}.
In our modeling approach, which assumes a linearly-increasing
velocity with radius, the more optically thick line forms further out, in a region that moves faster,
thereby showing a P~Cygni profile absorption further to the blue from line center. In other words,
the velocity shift of the absorption minimum amongst the Balmer lines suggests the velocity increases outwards
from the photosphere. This is different from the narrow absorption (and sometimes
associated narrow emission) that is sometimes seen on top of the strong and broad H$\alpha$ line
profile and that is associated with the CS material exclusively (see Kotak et al. 2004 for the
Type Ia SN 2002ic, Smith et al. 2007 for SN 2006gy, and Salamanca et al. 2002 for SN 1997eg).

\section{Discussion and conclusions}
\label{sect_discussion}

In this work, we have performed a quantitative spectroscopic analysis of the interacting
Type IIn SN 1994W, an unusual Type II event that exhibited narrow line cores with broad line wings in the optical,
was unusually luminous, showed erratic spectral behavior prior to peak brightness,
and synthesized an extremely low amount of $^{56}$Ni.
Our study covers specifically from 10-20 days prior to peak until 50 days afterwards, and makes
use of a radiative-transfer modeling approach that is one dimensional and steady-state,
assumes a (steep) power-law density distribution and a linear velocity law.
Importantly, it incorporates non-LTE effects and accurately treats the electron-scattering
source function.

During the brightening phase, also characterized by strong spectral variability,
we suspect the presence of multiple radiating regions,
split between an optically-thick layer contributing to the bulk of the optical light,
and shocked (and perhaps clumpy) material above contributing a smaller and rapidly-variable fraction.
This fraction is subdominant in the optical but could be much larger in the UV and X-ray ranges.
   Because this outer, low-density, material has been shocked to high temperatures, it
is optically thin and, thus, radiatively less efficient (see Fransson et al. 1996).
Such a dichotomy is also supported by the {\it simultaneous} presence of optical spectral
features testifying for both relatively low and high ionization conditions (i.e., Fe{\sc ii} and He{\sc i}).
Our models during this brightening phase are mostly exploratory, but suggestive of a complex
interaction configuration. Observations over a broad spectral band, and extending into at least the
near UV, are crucial for understanding these earlier phases.

\begin{figure}
\epsfig{file=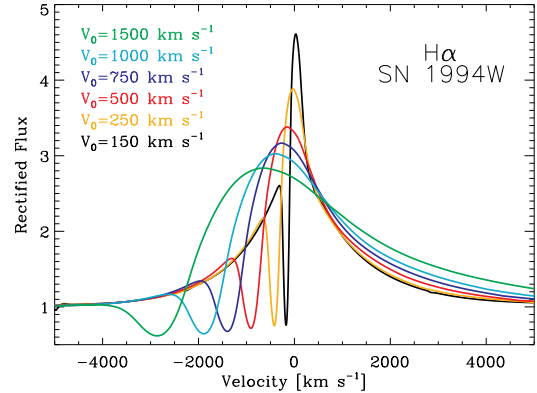,width=8cm}
\caption{Comparison between synthetic H$\alpha$ line profiles versus Doppler velocity obtained
with the model used to fit observations on 1994 October 11, but differing in the adopted
values of the base velocity $V_0$ (the velocity law is of the form $v(R) = V_0 (R/R_0){^\beta}$).
By decreasing $V_0$ from 1500\,\kms to 150\,\kms, we go from a broad P-Cygni profile typical
for a Type II-P SN to a symmetric line profile with a narrow line core and broad wings typical
for a Type IIn SN. (See text for discussion.)
\label{fig_comp_vcore}
}
\end{figure}

During the monotonic and slow fading phase, our fits are by contrast very satisfactory.
We find that the bolometric light must emerge entirely from a
single hydrogen-rich optically-thick layer moving at a near-constant velocity of $\sim$800\,\kms.
This photosphere recedes in both mass and radius with time, its extent shrinking from 4.3$\times$10$^{15}$
at peak to 2.3$\times$10$^{15}$\,cm 50 days later, while cooling modestly from
$\sim$7300 to $\sim$6300\,K over that same period.
As in Type II-P SNe, we find  that the photosphere is slaved to the region of full ionization,
which, as the material expands and radiates, can only shrink. The steep
fading 50 days after the peak is, thus, naturally associated with the emitting material
becoming completely optically thin.  The near-constant photospheric velocity of $\sim$800\,\kms,
for an extended period (over 60 days), places strong constraints on dynamical models.

 The above parameters result from the radiative transfer modeling of the
spectroscopic and photometric evolution of SN 1994W, and represent what is required
to reproduce observations. These properties should therefore be matched by any radiation hydrodynamics
modeling of SN 1994W - they represent an important guide.
In the context of an interaction between inner and outer ejecta, this photosphere
would likely reside somewhere between the reverse shock (located in the inner ejecta) and the forward shock
(located in the outer ejecta). The inferred photospheric velocity cannot be associated with CSM alone
since a large over-density and full-ionization is required to create this optically-thick layer,
a condition that is met only between the reverse and forward shocks (it cannot reside inside of the
reverse shock since this would be the inner fast ejecta and all emission profiles would appear
much broader than they do).

As this was a preliminary study, each epoch was modeled independently --- we did
not attempt to get consistency in the parameters between
different dates (i.e., the photospheric density at day 90 is lower
than what would be inferred by a power law extrapolation of the
photospheric density at day 50).  Whether a consistent density/velocity
set could be obtained will require a more consistent approach, possibly
in conjunction with dynamical modeling.  The results of such an
investigation would provide crucial constraints on the
dynamical model of the emitting region.

By contrast with C04, we do not find that the CDS represents such a key landmark for the understanding
of the light coming from SN 1994W.
The CDS may be optically thick at early times, but it must be optically thin after the peak to explain
the fading phase of the SN, which would otherwise be a brightening phase given the small photospheric
temperature decrease. A corollary is that the steep fading at late times corresponds to the onset
of the nebular phase rather than to the CDS leaving the external shell.

We reproduce successfully the narrow-core broad-wing line profile
morphology in the optical spectra of SN 1994W at all epochs, thus downplaying
the alleged role of an optically-thick external shell in causing this profile shape.
Instead, we find that this hybrid morphology results from the conditions at the
photosphere alone. Indeed, we argue that the small photospheric velocity of SN 1994W at all times
recorded makes electron scattering a key escape mechanism for line photons. The
associated frequency kicks redistribute photons from the optically-thick line cores
where they are trapped into the more optically-thin line wings, from where they can escape
after several scatterings with free electrons. The resulting flux in the wings increases
at the expense of the flux in the line core. Paradoxically, the slower the photospheric
velocity, the broader and more evident the line wings, which thus reflect the slow rather than the
fast expansion of the flow. As C04, we find that the effect is more pronounced for high
electron-scattering optical depth, but we associate this effect with the photosphere exclusively,
where both lines and continuum form, rather than with an external shell.
C04 estimated the properties of the external shell based on estimates of its
electron-scattering optical depth through the effect on line profiles. This needs
revision since the observed narrow line cores and broad line wings can be understood
from multiple-scattering internal to the photosphere.  Indeed, we can explain the entire
optical spectrum by a single emitting region. To avoid any misunderstanding, let us stress again. We are not proposing that there
is no external shell.  Rather, we find that the multiple electron-scattering events
at the origin of the broad line wings do not occur
in the external shell, but are instead internal to that localized, optically-thick, layer where most
photons are produced and which, we demonstrated, exists unambiguously.



\begin{figure*}
\epsfig{file=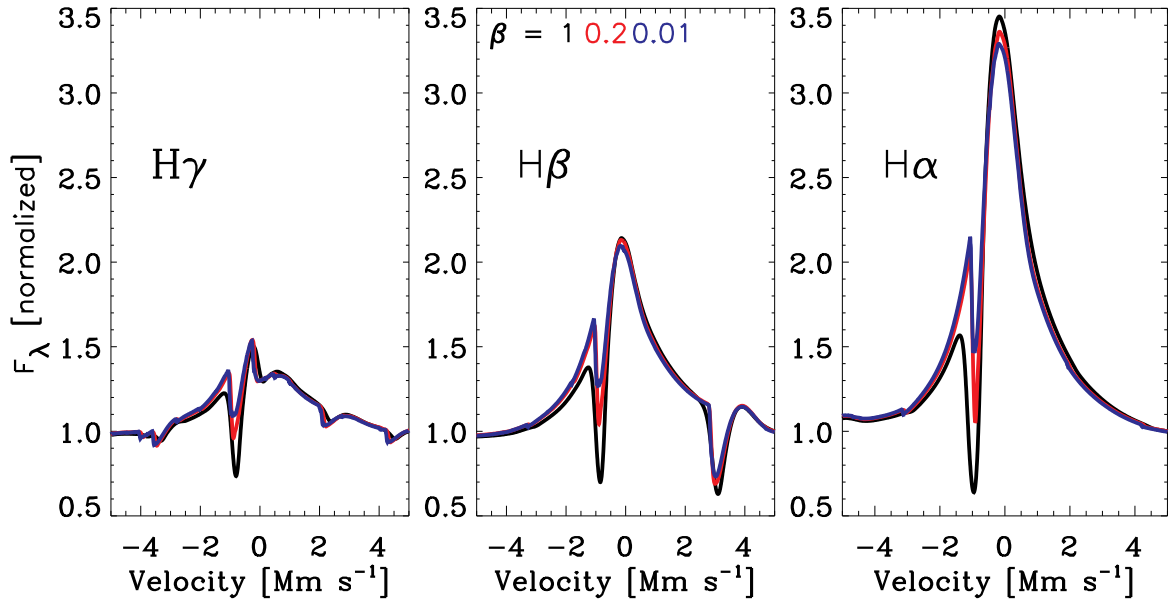,width=18cm}
\caption{
{\it Left:} Variation of the H$\gamma$ line profile computed for different assumptions
on the ejecta velocity distribution, i.e. characterized by $v(R) = V_0 (R/R_0){^\beta}$, with
$\beta$ equal to 1 (black; $V_{\rm Max}=$4000\,\kms), 0.2 (red; $V_{\rm Max}=$1364\,\kms), and 0.01
(blue; $V_{\rm Max}=$1123\,\kms). {\it Middle:} Same as left, but this time for H$\beta$.
{\it Right:} Same as left, but this time for H$\alpha$.
The reference model with $\beta=1$ is that
used to fit observations of SN 1994W on 1994 September 1 (see Fig.~\ref{fig_94W_0901}
and Table~1). For the latter two models, the base velocity is adjusted so that the photospheric
velocity is within 10\% of 830\,\kms, thence, producing a similar line core width.
Expansion causes little red-wing broadening since these three models, which differ so much in
maximum ejecta velocity $V_{\rm Max}$, show a similar red-wing extent.
Effects of non-coherent and multiple scattering by thermal electrons
is the primary cause of the extended wings. Note how the profile becomes more symmetric,
i.e. shows a stronger blue electron-scattering wing, as $\beta$ (or the velocity gradient,
which controls the escape probability of a photon trapped in the line) is
reduced.\label{fig_comp_flux_beta}
}
\end{figure*}

Let us now {\it speculate} on what may be at the origin of the SN 1994W event.
This speculation is motivated by two recent observations that show that Type IIn SNe are associated with widely different progenitors. First, objects like the Type IIn SN 2006gy, are
associated with a large bolometric luminosity (a rate in excess of 10$^{10}$\,\lsun\ sustained for months),
a large kinetic energy (there is interaction, hence deceleration, but still the line profiles are much broader than for SN 1994W),
and a large amount of mass in both the inner and outer ejecta that interact (because the ejecta remain optically thick
for months). These objects must be associated with very massive stars, losing considerable amounts
of mass in at least two events, and exploding with a prodigious energy which seems to require either
gravitational collapse or thermonuclear burning. Second, the Type IIn SN 2008S progenitor is identified on pre-explosion images as a star of $\sim$10\,\mo\ (Prieto et al. 2008). That two SNe with the same type be associated with
progenitors of such different properties suggests that such interactions can occur in a wide variety of
circumstances. Further, the observations (e.g., SN 2008S; Prieto et al. 2008) show that SN IIn are not necessarily associated with
LBVs.  Note that stars in the mass range 8-10\,\mo\
are considerably more numerous than high mass progenitors at the origin of SN 2006gy-like events.
Moreover, they do not build degenerate Fe cores, but ONeMg ones.

In the speculative statements we present below,
we wish to raise the issue that
1) given an interaction has to occur to explain the combination
of large photospheric radii and small expansion velocities (inferred from line profile widths), and
2) given that Type II SNe have typically 100 times more kinetic energy than radiant energy
(1\,B compared to 0.01\,B; 1B$\equiv$10$^{51}$\,erg), conversion efficiencies of a few
to a few tens of percent from kinetic to internal energy can satisfy the energetics of
most Type IIn SNe.
With such a huge kinetic energy reservoir, any modest conversion to internal energy can give rise
to SN-like displays. And indeed, the fact that statistically, the brightest SNe known are of Type IIn is compelling.
Hence, the kinetic energy and mass budget, the conversion efficiency from kinetic to internal energy,
and the expected variety of circumstellar material configurations, offer a natural way to explain
both the brightness and the diversity of Type IIn SNe (see introduction for a short synopsis of this diversity).
Our speculation, now, is that one may not systematically need 1\,B of kinetic energy to power a Type IIn SN
bolometric display.

Turning to SN 1994W, and adopting a representative
luminosity of 2$\times$10$^9$\,\lsun\, over a 100-day period, the time-integrated bolometric light
for SN 1994W is $E_{\rm rad} \approx 0.066$\,B. With a conversion efficiency of 30\% 
(which is at the upper end of what Chevalier (1983) proposes for driven waves;
variations of a few percent can be accommodated given the speculative level
meant in the present discussion), this would require
a kinetic energy of $E_{\rm kin} \approx 0.2$\,B, which can be matched by, for example,
a 2\,\mo\ shell moving at 3000\,\kms. One can try different combinations of mass and velocity, but the point here is
that this does not represent so much energy. This may not be what happens in SN 19994W, but it demonstrates
that modest kinetic energies, i.e. less than standard core-collapse SN explosion energies,
do not violate any of the energetics to power the bolometric displays like that of SN 1994W. Whether they
do in practice needs to be demonstrated.
In this context, it seems that a shell-shell interaction could explain in principle the radiation budget
of SN 1994W, and that a powerful core-collapse SN explosion may not be needed.
Additionally, the absence of core collapse and potential nickel production would satisfy the
very low nickel yields inferred for this SN, a record low of 0.0026--0.015\,\mo\ (SCL).
It also can more easily explain the absence of broad lines at all times.
The mechanism for the ejection of the second shell would also be the same
one as for the first shell, but the second ejection would have to expand faster to catch up the first.
We note that no
massive star is known to have a radius larger than 10$^{14}$\,cm (RSG, see Levesque et al. 2005),
so there should be ample time to detect and thoroughly observe the expansion of the inner shell prior to
interaction, in fact about 100/$V_3$ days (assuming constant velocity expansion), and determine its properties.
Note however that  none of these observations alone are sufficient to rule out a core collapse, but they do suggest that
other mechanisms should also be examined.
For example, a low Ni yield, by itself, does not rule out a core collapse SN.
Recent theoretical simulations of 8-10\,\mo\ stars indicate that the resulting SN can give low Ni yields
($<0.015\,$\mo)(e.g., Kitaura et al. 2006). SN 2005cs is regarded as a low luminosity SN,
is considered to have arisen from a  low mass progenitor (e.g., Eldridge et~al 2007),
and had a low Ni yield of $\sim 0.01\,$\mo\ (Pastorello et~al. 2006; Tsvetkov et al. 2006).
 Fall-back may also truncate the inner, nickel-rich ejecta (Woosley \& Timmes 1996), but the quantitative
aspects of this  process are uncertain, and one may wonder why these Type IIn SNe,
whose properties are really set by external interaction, would also have experienced significant fallback.




C04 inferred  the presence of a 0.4\,\mo\ CS envelope, while we know that core-collapse SN ejecta
have at least 6\,\mo, typically endowed with 1\,B kinetic energy.\footnote{Only stars more
massive than 8\,\mo\ on the  main sequence undergo core collapse, so, leaving aside 2\,\mo\ for
the neutron star and fallback material leaves us with at least 6\,\mo. More generic values
may be 10-15\,\mo.}
Given this unequal mass configuration, it is  unlikely the SN ejecta could be decelerated
throughout to velocities below 1000\,\kms. A more favorable configuration for deceleration,
supported by the narrow lines at all times, is for two shells of comparable mass. The mechanism causing the shell ejections is unknown, but may be related to pulsations or
nuclear flashes in the last stages of core burning. For example, within $\sim$2\,yr of core collapse,
20\,\mo\ main-sequence stars go through a $\sim$1.5\,yr-long oxygen core-burning phase: Instabilities associated
with that phase that led to an explosive shell ejection would provide a reproducible time delay before collapse
and an attractive mechanism at the origin of some Type IIn SNe like 1994W.
Woosley et al. (2007) have invoked the same mechanism,
the collision between ejected shells, to explain the extremely luminous supernova SN 2006gy.
In the model of Woosley et al. (2007) pair-instability pulsations in massive stars (e.g,
$M \sgreat 110\,$\mo) eject multiple shells which later collide.



Circumstellar envelopes with kinetic energies with $>$0.01\,B ergs are known.
In the 1840's, Eta Carinae underwent a major outburst ejecting over 10\,\mo\ of material
at a velocity of ~650\kms, with kinetic energy $\sgreat$0.04\,B (Smith et al. 2003).
Unfortunately, the duration of the ejection event is not well constrained, and
there is no consensus as to the cause of the event. More interestingly, Eta Carinae underwent a second
eruption, most likely in the 1890's, although the event was much less significant
(0.1\,\mo, $V\sim200$\,\kms,
10$^{-4}$\,B; Smith 2003). While the dynamics and kinematics don't satisfy the requirements for a bright
circumstellar interaction, they are in the least suggestive that such an interaction might occur.
A potential issue with shell ejections from LBVs and similar stars is that the ejection velocities
tend to scale with the escape speed. For SN 1994W, it is likely the second shell needed speeds of
order 1500\,\kms, or larger. A direct corollary of the shell interaction scenario is that the
progenitor star would not be destroyed (although it might be difficult to detect in the optical if
dust formed), and this offers a direct means of distinguishing between the core-collapse and
multiple shell-ejection scenarios. Similarly,
direct confirmation of Ni/Co decay could be used to distinguish between the two models.

Type IIn SNe gather a very heterogeneous group and our findings should not be applied blindly
to other Type IIn SNe. In some cases (e.g,  SN 1988Z, Stathakis  \& Sadler 1991; van Dyk et al. 1993;
Turatto et al. 1993; Chugai \& Danziger 1994) it is readily apparent that core-collapse SN ejecta,
interacting with circumstellar matter, can explain the observed spectral variations. To make further
progress, detailed radiation hydrodynamics of the interaction and detailed radiative transfer
calculations of emergent spectra need to be performed in partnership, and for a wide range of
configurations.  Allowance for time-dependent effects and departures from sphericity may also
be needed. Observations in the blue (as far to the UV as
possible), red, and near IR would provide critical constraints on the models.  As a key probe
of shocks, X-ray observations are crucial. Finally, public availability of all observational
data would help reveal fully the diversity of Type IIn spectra and light curves, thereby helping
understand this challenging group of SNe.



\section*{Acknowledgments}

We thank Adam Burrows, Rubina Kotak, Jeremiah Murphy, and Steve Smartt
for discussion,
Robert Cumming for providing the SN 1994W spectroscopic data used here,
Nikolai Chugai for his comments,
and St\'ephane Blondin for a thorough reading of a draft of this paper and for his suggestions.
L.D. acknowledges support for this work from the Scientific Discovery
through Advanced Computing (SciDAC) program of the DOE, under grant numbers
DOE-FC02-01ER41184 and DOE-FC02-06ER41452, and from the NSF under grant
number AST-0504947.


\appendix
\section{}

The following tables provide a summary of the spectral lines important for fitting SN 1994W at different
dates. Line EWs were computed assuming isolated lines, and using a modified form of the Sobolev approximation.
They are meant to provide a guide to the relative importance of various features in the spectrum, and should NOT be
used for quantitative analysis. Level name notation is based on the notation of levels in CMFGEN which in turn is based
on the notation from  NIST (Ralchenko et al.  2008).

\clearpage


\onecolumn

\begin{center}
\begin{longtable}{lrll}
\caption{Line list for model used to match observations on 1994 September 1} \label{tab_linelist} \\

\hline \hline \\[-2ex]
 $\lambda$ (\,\AA) & $W_{\lambda}$ (\,\AA)   & Species & Transition (upper level -- lower level) \\[0.5ex] \hline
   \\[-1.8ex]
\endfirsthead

\multicolumn{4}{c}{{\tablename} \thetable{} -- Continued} \\[0.5ex]
\hline \hline \\[-2ex]
 $\lambda$ (\,\AA) & $W_{\lambda}$(\,\AA)   & Species & Transition (upper level -- lower level) \\[0.5ex] \hline
   \\[-1.8ex]
\endhead

\\[0.5ex] \hline
   \\[-1.8ex]
  \multicolumn{4}{l}{{Continued on Next Page\ldots}}
\endfoot

  \\[-1.8ex] \hline \hline
\endlastfoot

 3213.3  &    -1.5 &  Fe{\,\sc ii}&  3d$^6$\,($^5$D)  4p\,\,z$^4$D$^{\rm o}_{5/2}$ -- 3d$^7$\,\,a$^4$P$_{3/2}$  \\
 3227.7  &    -1.6 &  Fe{\,\sc ii}&  3d$^6$\,($^5$D)  4p\,\,z$^4$D$^{\rm o}_{7/2}$ -- 3d$^7$\,\,a$^4$P$_{5/2}$  \\
 3662.3  &     1.2 &  H{\,\sc i}&  30 -- 2  \\
 3663.4  &     1.4 &  H{\,\sc i}&  29 -- 2  \\
 3664.7  &     1.5 &  H{\,\sc i}&  28 -- 2  \\
 3666.1  &     1.7 &  H{\,\sc i}&  27 -- 2  \\
 3667.7  &     1.9 &  H{\,\sc i}&  26 -- 2  \\
 3669.5  &     2.2 &  H{\,\sc i}&  25 -- 2  \\
\end{longtable}
\end{center}

\clearpage


\begin{longtable}{lrll}
\caption{Line list for model used to match observations on 1994 September 9} \label{tab_linelist} \\

\hline \hline \\[-2ex]
 $\lambda$ (\,\AA) & $W_{\lambda}$ (\,\AA)   & Species & Transition (upper level -- lower level) \\[0.5ex] \hline
   \\[-1.8ex]
\endfirsthead

\multicolumn{4}{c}{{\tablename} \thetable{} -- Continued} \\[0.5ex]
\hline \hline \\[-2ex]
 $\lambda$ (\,\AA) & $W_{\lambda}$(\,\AA)   & Species & Transition (upper level -- lower level) \\[0.5ex] \hline
   \\[-1.8ex]
\endhead

\\[0.5ex] \hline
   \\[-1.8ex]
  \multicolumn{4}{l}{{Continued on Next Page\ldots}}
\endfoot

  \\[-1.8ex] \hline \hline
\endlastfoot

 3210.4  &    -1.4 &  Fe{\,\sc ii}&  3d$^6$\,($^5$D)  4p\,\,z$^4$D$^{\rm o}_{3/2}$ -- 3d$^7$\,\,a$^4$P$_{1/2}$  \\
 3213.3  &    -2.1 &  Fe{\,\sc ii}&  3d$^6$\,($^5$D)  4p\,\,z$^4$D$^{\rm o}_{5/2}$ -- 3d$^7$\,\,a$^4$P$_{3/2}$  \\
 3227.7  &    -2.2 &  Fe{\,\sc ii}&  3d$^6$\,($^5$D)  4p\,\,z$^4$D$^{\rm o}_{7/2}$ -- 3d$^7$\,\,a$^4$P$_{5/2}$  \\
 3408.8  &    -1.2 &  Cr{\,\sc ii}&  3d$^4$\,($^5$D)  4p\,\,z$^6$D$^{\rm o}_{5/2}$ -- 3d$^4$\,($^5$D)  4s\,\,a$^4$D$_{7/2}$  \\
 3422.7  &    -1.0 &  Cr{\,\sc ii}&  3d$^4$\,($^5$D)  4p\,\,z$^4$P$^{\rm o}_{3/2}$ -- 3d$^4$\,($^5$D)  4s\,\,a$^4$D$_{5/2}$  \\
 3442.0  &    -1.1 &  Mn{\,\sc ii}&  3d$^5$\,($^6$S)  4p\,\,z$^5$P$^{\rm o}_{3}$ -- 3d$^6$\,\,a$^5$D$_{4}$  \\
 3514.0  &    -1.3 &  Ni{\,\sc ii}&  3d$^8$\,($^3$F) 4p$\,\,^4$D$^{\rm o}_{7/2}$ -- 3d$^8$\,($^3$P)  4s$\,\,^4$P$_{5/2}$  \\
 3662.3  &     1.2 &  H{\,\sc i}&  30 -- 2  \\
\end{longtable}

\clearpage


\begin{longtable}{lrll}
\caption{Line list for model used to match observations on 1994 October 1} \label{tab_linelist} \\

\hline \hline \\[-2ex]
 $\lambda$ (\,\AA) & $W_{\lambda}$ (\,\AA)   & Species & Transition (upper level -- lower level) \\[0.5ex] \hline
   \\[-1.8ex]
\endfirsthead

\multicolumn{4}{c}{{\tablename} \thetable{} -- Continued} \\[0.5ex]
\hline \hline \\[-2ex]
 $\lambda$ (\,\AA) & $W_{\lambda}$(\,\AA)   & Species & Transition (upper level -- lower level) \\[0.5ex] \hline
   \\[-1.8ex]
\endhead

\\[0.5ex] \hline
   \\[-1.8ex]
  \multicolumn{4}{l}{{Continued on Next Page\ldots}}
\endfoot

  \\[-1.8ex] \hline \hline
\endlastfoot

 3202.5  &    -3.4 &  Ti{\,\sc ii}&  3d$^2$\,($^1$D)  4p\,\,y$^2$F$^{\rm o}_{5/2}$ -- 3d$^2$\,($^1$D)  4s\,\,a$^2$D$_{3/2}$  \\
 3203.4  &    -1.4 &  Ti{\,\sc ii}&  3d$^2$\,($^3$F)  4p\,\,z$^2$F$^{\rm o}_{5/2}$ -- 3d$^2$\,($^3$F)  4s\,\,a$^4$F$_{3/2}$  \\
 3210.4  &    -5.1 &  Fe{\,\sc ii}&  3d$^6$\,($^5$D) 4p\,\,z$^4$D$^{\rm o}_{3/2}$ -- 3d$^7$\,\,a$^4$P$_{1/2}$  \\
 3211.1  &    -1.1 &  Fe{\,\sc ii}&  3d$^6$\,($^3$F2) 4p\,\,x$^4$D$^{\rm o}_{7/2}$ -- 3d$^7$\,\,b$^2$F$_{5/2}$  \\
 3212.0  &    -2.7 &  Fe{\,\sc ii}&  3d$^6$\,($^3$H)  4p\,\,z$^4$I$^{\rm o}_{9/2}$ -- 3d$^6$\,($^3$G)  4s\,\,b$^2$G$_{9/2}$  \\
 3213.3  &    -6.4 &  Fe{\,\sc ii}&  3d$^6$\,($^5$D) 4p\,\,z$^4$D$^{\rm o}_{5/2}$ -- 3d$^7$\,\,a$^4$P$_{3/2}$  \\
 3214.8  &    -2.0 &  Ti{\,\sc ii}&  3d$^2$\,($^3$F)  4p\,\,z$^2$F$^{\rm o}_{7/2}$ -- 3d$^2$\,($^3$F)  4s\,\,a$^4$F$_{9/2}$  \\
 3217.1  &    -4.1 &  Ti{\,\sc ii}&  3d$^2$\,($^3$F)  4p\,\,z$^4$F$^{\rm o}_{9/2}$ -- 3d$^2$\,($^3$F)  4s\,\,a$^4$F$_{7/2}$  \\
\end{longtable}

\clearpage


\begin{longtable}{lrll}
\caption{Line list for model used to match observations on 1994 October 11} \label{tab_linelist} \\

\hline \hline \\[-2ex]
 $\lambda$ (\,\AA) & $W_{\lambda}$ (\,\AA)   & Species & Transition (upper level -- lower level) \\[0.5ex] \hline
   \\[-1.8ex]
\endfirsthead

\multicolumn{4}{c}{{\tablename} \thetable{} -- Continued} \\[0.5ex]
\hline \hline \\[-2ex]
 $\lambda$ (\,\AA) & $W_{\lambda}$(\,\AA)   & Species & Transition (upper level -- lower level) \\[0.5ex] \hline
   \\[-1.8ex]
\endhead

\\[0.5ex] \hline
   \\[-1.8ex]
  \multicolumn{4}{l}{{Continued on Next Page\ldots}}
\endfoot

  \\[-1.8ex] \hline \hline
\endlastfoot

 3202.5  &    -4.0 &  Ti{\,\sc ii}&  3d$^2$\,($^1$D)  4p\,\,y$^2$F$^{\rm o}_{5/2}$ -- 3d$^2$\,($^1$D)  4s\,\,a$^2$D$_{3/2}$  \\
 3203.4  &    -2.7 &  Ti{\,\sc ii}&  3d$^2$\,($^3$F)  4p\,\,z$^2$F$^{\rm o}_{5/2}$ -- 3d$^2$\,($^3$F)  4s\,\,a$^4$F$_{3/2}$  \\
 3210.4  &    -4.8 &  Fe{\,\sc ii}&  3d$^6$\,($^5$D) 4p\,\,z$^4$D$^{\rm o}_{3/2}$ -- 3d$^7$\,\,a$^4$P$_{1/2}$  \\
 3212.0  &    -2.3 &  Fe{\,\sc ii}&  3d$^6$\,($^3$H)  4p\,\,z4I$^{\rm o}_{9/2}$ -- 3d$^6$\,($^3$G)  4s\,\,b$^2$G$_{9/2}$  \\
 3213.3  &    -6.2 &  Fe{\,\sc ii}&  3d$^6$\,($^5$D) 4p\,\,z$^4$D$^{\rm o}_{5/2}$ -- 3d$^7$\,\,a$^4$P$_{3/2}$  \\
 3214.8  &    -3.2 &  Ti{\,\sc ii}&  3d$^2$\,($^3$F)  4p\,\,z$^2$F$^{\rm o}_{7/2}$ -- 3d$^2$\,($^3$F)  4s\,\,a$^4$F$_{9/2}$  \\
 3216.9  &    -1.5 &  Ti{\,\sc ii}&  3d$^2$\,($^1$D)  4p\,\,y$^2$F$^{\rm o}_{7/2}$ -- 3d$^3$ a$^2$G$_{7/2}$  \\
 3217.1  &    -5.5 &  Ti{\,\sc ii}&  3d$^2$\,($^3$F)  4p\,\,z$^4$F$^{\rm o}_{9/2}$ -- 3d$^2$\,($^3$F)  4s\,\,a$^4$F$_{7/2}$  \\
\end{longtable}

\twocolumn


\label{lastpage}

\end{document}